\documentclass{aa}
\usepackage{graphicx}
\usepackage{caption}
\usepackage{txfonts}
\usepackage{epstopdf}
\usepackage{color}
\usepackage{multirow}
\usepackage{lscape}
\usepackage{amsmath}
\usepackage{amssymb}
\usepackage{upgreek}
\usepackage{hyperref}

\usepackage{subcaption}
\usepackage{comment}
\usepackage{upgreek}
\usepackage[dvipsnames]{xcolor}

\begin{document}

\title{Modelling of eclipsing binary systems with pulsating components and tertiary companions: BF~Vel and RR~Lep}
     \author{A. Liakos\inst{1}
             \and
              D. J. W. Moriarty\inst{2,3}
             \and
              A. Erdem\inst{4,5}
             \and
              J. F. West\inst{3}
             \and
             P. Evans\inst{6} }

   \institute{Institute for Astronomy, Astrophysics, Space Applications and Remote Sensing, National Observatory of Athens,\\
              Metaxa \& Vas. Pavlou St., GR-15236, Penteli, Athens, Greece \\
              \email{alliakos@noa.gr}
              \and
              School of Mathematics and Physics, The University of Queensland, Queensland 4072, Australia
              \and
              Astronomical Association of Queensland, St. Lucia, Queensland, 4067, Australia
             \and
              Astrophysics Research Center and Ulup{\i}nar Observatory, \c{C}anakkale Onsekiz Mart University, TR-17100, \c{C}anakkale, T\"{u}rkiye
              \and
              Department of Physics, Faculty of Arts and Sciences, \c{C}anakkale Onsekiz Mart University, Terzio\u{g}lu Kamp\"{u}s\"{u}, TR-17100, \\ \c{C}anakkale, T\"{u}rkiye
              \and
              El Sauce Observatory, Coquimbo Province, Chile
              }
           \date{Received XX January 2024; accepted XX January 2024}


\abstract
{This research paper presents a comprehensive analysis of RR~Lep and BF~Vel, two southern short-period semi-detached oscillating Algols (oEA stars), which are shown to be triple systems. Spectral types of the primary components were determined and radial velocities calculated from spectra observed with the Australian National University's 2.3~m telescope and Wide Field Spectrograph. Spectra of the Na~I~D doublet confirmed the presence of tertiary components which were apparent in the broadening function analyses and, with H$_\alpha$ spectra during primary eclipses, indicated chromospherical activity in their secondary components. Ground-based telescopes were used for observations in several pass bands for photometric analyses. These observations were complemented by data from the Transiting Exoplanet Survey Satellite (TESS) mission to enable the modelling of the light curves, followed by a detailed analysis of pulsations. Eclipse-timing variation (ETV) analyses of both systems were used to determine the most likely mechanisms modulating the orbital period. We found mass values $M_1=2.9$~M$_{\sun}$ and $M_2=0.75$~M$_{\sun}$ for the components of RR~Lep, and $M_1=1.93$~M$_{\sun}$ and $M_2=0.97$~M$_{\sun}$ for those of BF~Vel. By integrating information from photometry, spectroscopy and ETV analysis, we found that tertiary components revolve around both systems. The primary star of RR~Lep pulsates in 36 frequencies, of which five were identified as independent modes, with the dominant one being 32.28~d$^{-1}$. The pulsating component of BF~Vel oscillates in 37 frequencies, with the frequency 46.73~d$^{-1}$ revealed as the only independent mode. For both systems, many frequencies were found to be related to the orbital frequency. Their physical properties were compared with other oEA stars in Mass-Radius and Hertzsprung-Russell diagrams, and the pulsational properties of their $\delta$~Sct components were compared with currently known systems of this type within the orbital-pulsation period and $\log g$-pulsation period diagrams.}

\keywords{binaries: eclipsing -- stars: fundamental parameters – binaries: close – stars: oscillations – stars: variables: delta Scuti -- stars: individual (BF~Vel) -- stars: individual (RR~Lep)}

\maketitle
%

\section{Introduction}
\label{Sec:Intro}

The $\delta$~Sct stars exhibit rapid and multi-periodic pulsations occurring within a frequency range of 4-80~d$^{-1}$. Their pulsation behaviour is predominantly driven by radial and low-order non-radial $p$-modes, triggered by the $\kappa$-mechanism \citep[cf.][and references therein]{EDD26, BAK62, ZHE63, BRE00, AER10, BAL15}. Furthermore, they also pulsate in high-order non-radial modes, influenced by turbulent pressure within the hydrogen convective zone \citep{ANT14}. These stars typically have masses ranging from 1.5-2.5~M$_{\sun}$ and belong to spectral classes between A and early F. They are primarily situated within the classical instability strip, spanning from the main-sequence to the giants branch \citep{AER10}.

Binary stars are stellar systems comprising two gravitationally bound components orbiting a common center of mass. They serve as crucial astrophysical tools for determining fundamental stellar characteristics, especially mass, radius, and temperature. Close binaries, characterized by orbital periods ranging from less than one day up to tens of days, have properties that potentially allow for mass transfer between the components \citep[cf.][and references therein]{KOP59, LOO92, KAL99, HIL01}. Eclipsing binaries (EBs) constitute a specific subset, where the orbital plane aligns with the observer's line of sight, resulting in mutual eclipses of the components. These eclipses induce periodic brightness variations, with their depth, shape, and duration offering insights into various orbital and physical properties such as inclination, temperature, radii, and mass function \citep[cf.][]{PRS18}. Furthermore, long-term monitoring of eclipses using the Eclipse-Timing Variation (ETV) technique reveals orbital period modulations attributable to physical mechanisms like mass transfer or loss. Spectroscopic binaries (SBs) exhibit detectable periodic Doppler shifts in their spectra due to the components' orbital motion around the common center of mass. These spectral line shifts enable determination of the components' radial velocities and hence their mass ratio. SBs are classified as single-lined (SB1) or double-lined (SB2) respectively, depending on whether absorption or emission lines from one or both components are detected. Notably, SB1(2)+EB systems are particularly significant as they provide direct measurements of both orbital and absolute physical parameters of the binary components.

\citet{MKR02} coined the term `oEA~stars' (oscillating EBs of Algol type) to describe eclipsing pairs with a $\delta$~Sct mass accretor component of spectral type (B)A-F. The impact of mass accretion on pulsations remains a key unresolved question in this branch of asteroseismology \citep[cf.][]{TKA09, BOW19, MKR22}. The association between orbital ($P_{\rm orb}$) and dominant pulsation ($P_{\rm pul}$) periods for binaries, including those with a $\delta$~Sct member, was first noted by \citet{SOY06a}. Following a comprehensive six-year survey of candidate systems, \citet{LIA12} refined this relationship with an expanded dataset. \citet{ZHA13} presented the initial theoretical endeavor to establish this connection. Further progress came with the work of \citet{LIAN17}, who compiled an updated catalogue containing approximately 200 known cases, from which only 20 semi-detached and 15 detached systems had precise absolute parameters. They also proposed the existence of subgroups exhibiting distinct properties based on the Roche geometry of their host systems and a boundary of 13-days in the orbital period, beyond that the pulsations are not affected by the companion's presence. \citet{KAH17}, focusing solely on EBs, suggested an approximately doubled threshold for $P_{\rm orb}$.

The exceptional precision (in the order of 10$^{-4}$~mag) and continuous time coverage of photometric data from the Transiting Exoplanet Survey Satellite mission \citep[TESS;][]{RIC15}  offer a valuable resource for in-depth investigations of stellar oscillations. This capability is particularly potent when coupled with multi-band photometry and high-resolution spectroscopy. The temporal resolutions of data collected by the TESS missions range from 20~seconds to 30~minutes, proving highly effective for identifying short-period frequencies, a characteristic feature of $\delta$~Sct stars. Notably, owing to their precision, they also enable the detection of low-amplitude pulsations at the level of a few $\upmu$mag \citep[cf.][]{KUR20, KIM21, LIA22}. Furthermore, these missions afford extensive temporal coverage, sometimes spanning years, which proves invaluable for determining the timing of minima in EBs. This, in turn, facilitates a more thorough investigation of their orbital period variations.

According to the catalogue of \citet{LIAN17} and the updated list hosted by AL (which will be published in a future work), there are currently more than 1000 known cases of binaries with $\delta$~Sct components. The vast majority (more than 90\%) of these systems lack information about the physical and (or) detailed pulsation parameters of their oscillating members. At present, there are approximately only 100 cases with well determined parameters (i.e. they are of the SB1(2)+EB types) and they are distributed almost equally into detached and semi-detached systems, with less than 50 cases per group. From these systems, only 33 oEA stars are SB2+EB systems. Therefore, new studies providing accurate results regarding the absolute parameters of the pulsating stars accompanied with high photometric data from space missions are extremely important for Asteroseismology.

This paper is a continuation of the detailed study on individual southern oEA stars with a $\delta$~Sct component. We report analyses of  two EBs, namely RR~Leporis and BF~Velorum. These systems are previously known EBs exhibiting pulsations (see next paragraphs for historical elements), but the absolute properties of their components have not been determined accurately to date. Therefore, since both of them were found to be SB2+EB systems, this work enriches the sample of oEAs with well-determined absolute stellar properties by 6\%.

RR~Leporis had been observed photoelectrically by \citet{BOO86}, \citet{ABH89}, \citet{VYA89}, and \citet{SAM89}. In the latter work there was a slight evidence for pulsational behaviour, which was confirmed later by the CCD observations of \citet{DVO09}. \citet{LIA13} using systematic multi-filter CCD observations, obtained full light curves from which they derived an orbital model solution and determined the dominant pulsation frequency as 33.3~d$^{-1}$.  The light curves of \citet{VYA89} were analysed by \citet{ERD16}, who found that orbital period changes of the system resulted in a downward parabolic behaviour superimposed by a periodic curve. The latter was attributed to the possible existence of a low-mass tertiary component. \citet{KAH24}, using \'{E}chelle spectra, reported the system to be a single lined binary and calculated the radial velocity curve of  the primary component; they calculated a photometric model of the system using TESS data. They derived mass values of 1.79~M$_{\sun}$ and 0.5~M$_{\sun}$ for the primary and secondary components of the system, respectively. Moreover, they found 14 pulsation frequencies, from which six were identified as independent modes.

The system BF~Velorum is a rather neglected system. The only available light curves come from \cite{MAN09}. In that work, the first photometric model of this system was presented and the pulsational behaviour of the primary component was described.

For these two systems we have undertaken extensive data collection, including medium resolution spectroscopic data (Sect.~\ref{Sec:Spectroscopy}), as well as multi-band ground-based and space-borne photometric observations (Sect.~\ref{Sec:Photometry}). We aimed to: a) model the orbital properties of the systems (Sect.~\ref{Sec:Models}); b)~Determine stellar parameters with high precision (Sect.~\ref{Sec:AbsPar}); c)~Perform Eclipse-Timing Variation (ETV) analysis to identify the orbital period modulating mechanisms (Sect.~\ref{Sec:ETV}); d)~Detect pulsation frequencies and estimate oscillation modes (Sect.~\ref{Sec:Puls}). The findings and conclusions derived from these analyses are discussed in Sect.~\ref{Sec:Disc}.

\begin{table*}		
\begin{center}																	
\caption{Log of spectroscopic observations with the wide field spectrograph on the ANU 2.3~m telescope.}																	
\label{table:SPLOG}		
\scalebox{0.95}{
\begin{tabular}{l cccc cccc}																	
\hline \hline																	
System	&	Dates	&	Orbital phase	&	Number of spectra	&	Exp. time	&	Grating	&	S/N	&	Grating	&	S/N	\\
	&	(DD/MM/YY)	&		&		&	(s)	&		&		&		&		\\						
\hline																	
\multirow{6}{*}{BF Vel}	&	14/2/17	&	0	&	3	&	60	&	B3000	&	137	&	R3000	&	90	\\
	&	14/2/17	&	0.99	&	3	&	90	&	B7000	&	114	&	R7000	&	110	\\
	&	11/4/17	&	0.5	&	3	&	120	&	B3000	&	194	&	R3000	&	177	\\
	&	11/4/17	&	0.5	&	3	&	180	&	B7000	&	162	&	R7000	&	156	\\
	&	14/2/17-1/11/20	&	0.2-0.34	&	11	&	105-420	&	B7000	&	124-428	&	 R7000	&	118-412	\\
	&	13/2/17-25/4/21	&	0.69-0.78	&	20	&	240-300	&	B7000	&	210-360	&	 R7000	&	200-348	\\
 \hline
\multirow{3}{*}{RR~Lep}	&	2/12/20	&	0.5	&	3	&	60	&	B3000	&	250	&	 R7000 	&	164	\\
	&	15/3/20-30/10/21	&	0.17–0.29	&	10	&	180-240	&	B7000	&	254-360	&	 R7000	&	254-360	\\
	&	1-3/11/20	&	0.65-0.79	&	13	&	120-420	&	B7000	&	240-440	&	 R7000	&	230-440	\\
\hline																	
\end{tabular}		}															
\end{center}																	
\end{table*}	

\begin{figure*}
\centering
\includegraphics[width=16cm]{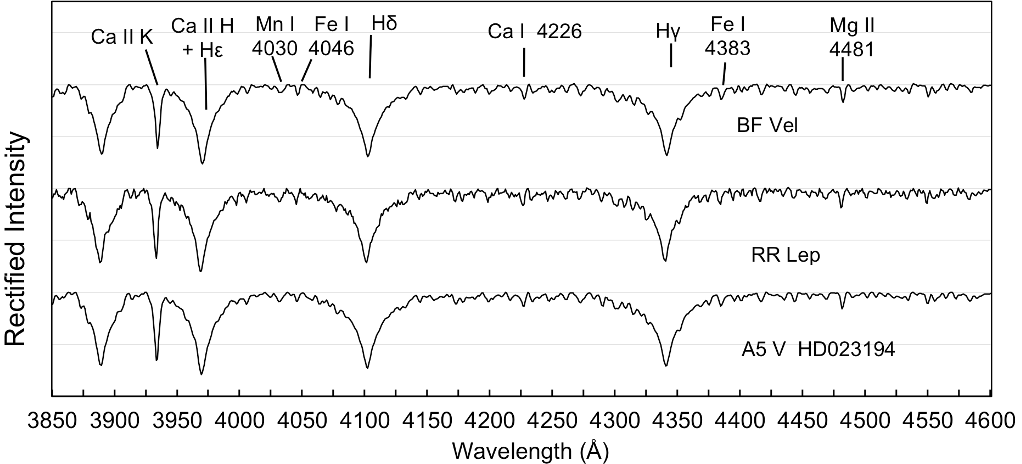}
\caption{Rectified spectra of the primary components of BF~Vel and RR~Lep and for comparison, the A5V star HD~23194.}
\label{Fig:spectra}
\end{figure*}

\section{Spectroscopy}
\label{Sec:Spectroscopy}

Spectra were observed with the stellar aperture on the wide field spectrograph (WiFeS) mounted at the Nasmyth focal position on the Australian National University's 2.3~m telescope at Siding Spring Observatory (SSO), Australia. The detailed observations log is given in Table~\ref{table:SPLOG}. A brief overview of the spectrograph is given here; for full details see \citet{DOP07,DOP10}. The 560~nm dichroic beam splitter was used. Fairchild Imaging CCD cameras with a resolution of $4096\times4096$~pixels and a pixel scale of 0.5~arcsec were used; one was optimised for the blue channel and the other for the red channel. The Ne-Ar arc lamp was used for wavelength calibration; the precision of the wavelength calibration was improved by observing arc spectra before and after sets of target spectra. Spectra were reduced with {\sc Pywifes}, a data reduction pipeline that was written for the WiFeS \citep{CHI14}.

One dimensional spectra were extracted from the 3D data cubes produced by the {\sc Pywifes} routines using a zero pixel radius around the spatial pixel where the target was centred and visualised in {\sc Qfitsview}\footnote{ http://www.mpe.mpg.de/~ott/dpuser/qfitsview.html}. This procedure was used also to avoid pixels with noise from probable cosmic ray hits that were not fully removed in the reduction routines. The sky background flux values were generally about two orders of magnitude lower than those of the target spectra. Flux units are ergs~cm$^{-2}$~s$^{-1}$~arcsec$^{-2}$~{\AA}$^{-1}$. See \citet{MOR19} and \citet{LIA22} for more details of the equipment and procedures used.

\subsection{Spectral classification}
\label{Sec:SpectrClass}

Spectra for classification were observed with the B3000 grating ($R=3000$) which has a wavelength range between 3200 and 5900~{\AA} that includes the important Ca~II~K line. The spectra of the primary components of BF~Vel and RR~Lep were observed with relatively short exposures during the secondary eclipse (phase 0.5) to minimise interference by their secondary and tertiary components.

Spectral types were determined using the {\sc Xclass} programme and the {\sc Mkclass}\footnote{http://www.appstate.edu/~grayro/mkclass} reference libraries \citep{GRA14}. {\sc Xclass} provides a direct visual comparison of the target spectrum with reference spectra from the {\sc Mkclass} libraries, which encompass integer spectral temperature types from O6 to M5, and a wide range of luminosity classes between V and Ia. The target spectra were normalised to unity at 4503~{\AA} to match the library spectra. The spectra were classified initially by visual comparisons with selected reference spectra across the wavelength range of 3800–-4600~{\AA}. In addition, careful reference was made to the structure of the Balmer lines, and to several individual metallic absorption lines, including Ca~II~K, Ca~I~4226, and Mg~II~4481. They were also compared to other A5V reference spectra, such as that of HD~23194, drawn from the MILES Library, and the examples provided in the work of \citet{GRA09}.

The best fit for the primary components of BF~Vel and RR~Lep was A5V. The spectra were rectified to deliver a linear continuum baseline using the ‘Autorectify’ function in the XMK25 display routine in {\sc Xclass} (see Fig.~\ref{Fig:spectra}).

\subsection{Radial velocities}
\label{Sec:RVs}

The B7000 and R7000 gratings were used to capture spectra for radial velocity determinations and for indications of chromospherical activity during the primary eclipse of BF~Vel. Exposure times with the B7000 grating were long enough to allow detection of the faint secondary stars' metal lines. At the same time as long exposure times were used with the B7000 grating, shorter exposure times were used with the R7000 grating to minimise blending and enhance the separation of Na~I~D lines of the primary star from those of the tertiary component in each system.

Radial velocities (RVs) of the systems and the binary components were calculated with the broadening function method implemented in the {\sc Ravespan} code written by B.~Pilecki \citep{RUC92, RUC02, PIL17} and are listed in Table~\ref{Tab:RVs}. Template spectra were taken from the synthetic spectra library of \citet{COE05}. The procedures are described in general by \citet{MOR19} and \citet{LIA22}, with {\sc Ravespan} settings chosen specifically for the systems in this paper as follows. The spectra with a wavelength range 4600--4840~{\AA} and 4880--5540~{\AA} were normalised in {\sc Ravespan}; a preference setting of velocity range $1.6 + 60$ was chosen as that gave the best fit of broadening function peak velocities to velocities observed in the Na~I~D spectra; other preference settings were the gravity coefficient 3.0, metallicity 0 and resolution 2~km~s$^{-1}$. The temperature settings of the synthetic spectrum templates were 5250~K for BF~Vel and 5000~K for RR~Lep. These settings excluded the Balmer lines and far blue spectral region to emphasise the lines of the late spectral type secondary component of each system.

\begin{figure}
\centering
\includegraphics[width=8.5cm]{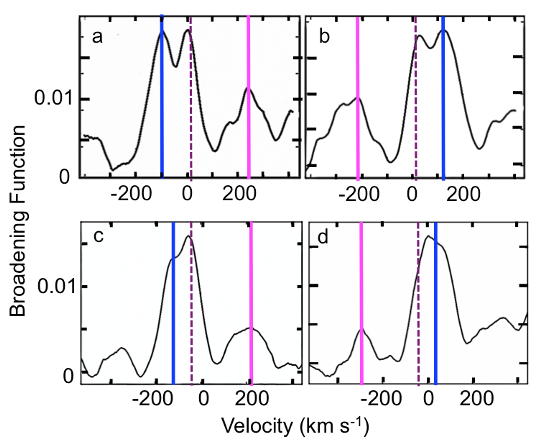}
\caption{Examples of broadening function curves. The orbital velocity of the primary components is marked with a blue line and that of the secondary components with a magenta line. The systemic velocity is indicated by the dashed line. (a)~BF~Vel phase 0.28, 2020 Nov.~1, exp.~time=240~s. (b)~BF~Vel phase 0.73, 2020 Nov.~2, exp.~time=240 s. (c)~RR~Lep phase 0.25, 2020 Dec.~1, exp.~time=240~s. (d)~RR~Lep phase 0.76, 2020 Nov.~3, exp.~time=420~s.}
\label{Fig:BFs}
\end{figure}

The presence of a tertiary star in each of these systems was evident in the broadening function curve of the primary component, either as a double peak with one peak having a low or zero velocity or as an asymmetrical curve (see Fig.~\ref{Fig:BFs}). The RVs of the secondary and tertiary components of BF~Vel were shown clearly at the quadrature phases in Fig.~\ref{Fig:BFs}a,~b. The RVs of the secondary component of RR~Lep were clearly evident in the broadening function analyses, whereas broadening function peaks of the primary and tertiary components of RR~Lep were merged (Fig.~\ref{Fig:BFs}c,~d). As described below, spectra of the Na~I~D doublet and H$_\alpha$ were also used to determine RVs and provide initial values of the primary components' velocities as inputs in {\sc Ravespan}.

\subsection{H$_\alpha$ spectra}
\label{Sec:Ha}

\begin{figure}[t]
\centering
\includegraphics[width=\columnwidth]{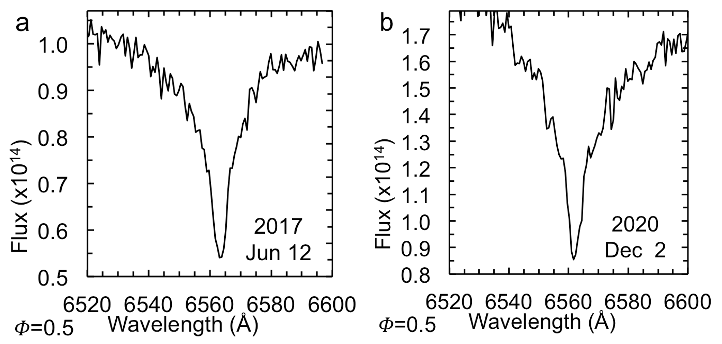}
\caption{H$_\alpha$ spectra at phase 0.5 (secondary eclipse). (a)~BF~Vel, exp.~time=240~s; (b)~RR~Lep, exp.~time=60~s.}
\label{Halpha}
\end{figure}

The systemic velocity of each system determined with radial velocity analyses was subject to uncertainty due to blending of the primary components' spectral lines with those of the tertiary components. As the spectral class of the primary components is A5V, their Balmer lines are much stronger than those of the secondary and tertiary components which have later spectral types, indicated by their sodium spectra (see Sect.~\ref{Sec:Na}). Therefore, spectra during secondary eclipses were observed to provide independent values for the systemic velocities (see Fig.~\ref{Halpha}). With relatively short exposures, the Balmer lines at phase 0.5 were dominated by those of the bright A5V primary stars, whereas during primary eclipses, the Balmer lines are a composite of the three components in each system. The H$_\alpha$ line centre of BF~Vel was at 6563.1~{\AA}, that is at 14~km~s$^{-1}$, which is a little above the range of $4.7\pm 1$~km~s$^{-1}$ determined by the modelling (Sect.~\ref{Sec:Models}). The H$_\alpha$ line is broadened by the combination of the primary component's rapid rotational velocity and blending with that of the tertiary star. The H$_\alpha$ line centre of RR~Lep was at 6561.7~{\AA}, indicating a systemic velocity of $\sim-50$~km~s$^{-1}$. The uncertainty of the velocities determined from the H$_\alpha$ spectra is about $\pm 5$~km~s$^{-1}$ due to the low-to-medium resolution of $R=7000$ of the wide field spectrograph and the broadening of the Balmer lines by the rapid rotational velocities of the primary components of the systems (Sect.~\ref{Sec:Models}). The H$_\alpha$ line of RR~Lep is broadened to the red due to the tertiary component.

\subsection{Sodium D doublet spectra}
\label{Sec:Na}

\begin{figure}
\centering
\includegraphics[width=\columnwidth]{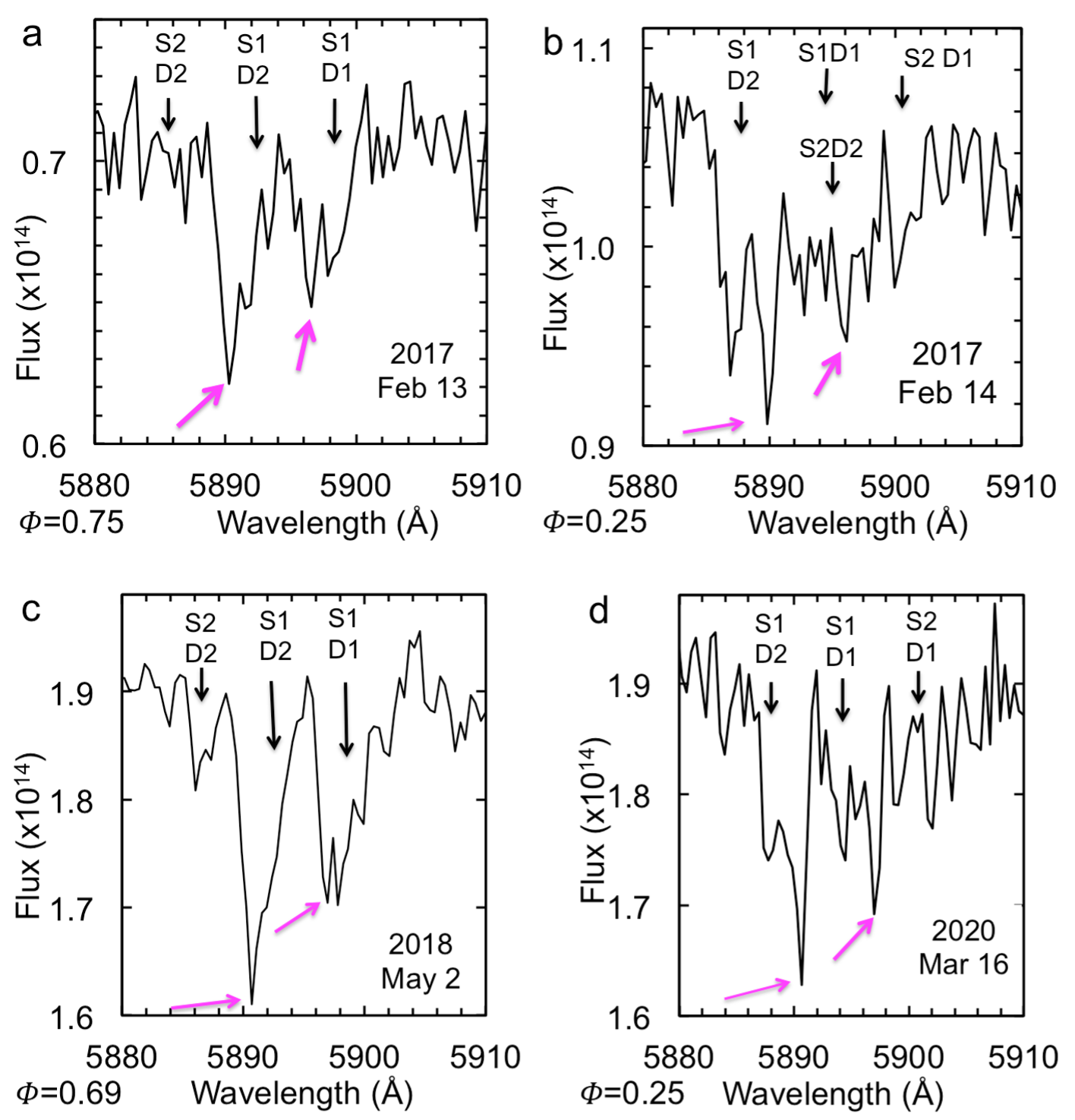}
\caption{Spectra of the BF~Vel Na~I~D doublet at orbital phases ($\phi$) around quadrature; $R=7000$. Downward arrows mark the calculated positions of the line centres for the primary star (S1) and secondary (S2) that give the best fit to the {\sc RAVESPAN} analyses. Lines of the tertiary component are indicated with magenta arrows angled up. Exposure times: (a)~300~s; (b)~180~s; (c)~300~s; (d)~105~s.}
\label{sodiumD_bfvel}
\end{figure}

Spectra of the Na~I~D doublet were used to check the velocities calculated in {\sc Ravespan}. At quadrature phases, the lines of the binary pairs are separated from those of the tertiary components to an extent dependent on their orbital velocities and length of the observation exposures. The sodium~D lines of the tertiary star in the BF~Vel system were apparent, although partially blended with those of the primary component, at phase 0.75 with an exposure time of 300~s (Fig.~\ref{sodiumD_bfvel}a). Note at that phase, the D2 line of the secondary component of BF~Vel was not distinguishable from noise. One night later, at the opposite quadrature and with a shorter exposure of 180~s, there was a distinct separation of the tertiary star's D2 line from that of the primary star; the tertiary D1 line was also noticeable, although partially blended with the lines of the primary and secondary components (Fig.~\ref{sodiumD_bfvel}b). Similarly, at phase 0.69 with a 300~s exposure, the narrow lines of the tertiary star were blended, but visible, and at phase 0.25 with a shorter exposure, were clearly separated from the primary component's sodium lines (Fig.~\ref{sodiumD_bfvel}c,~d). The D2 line of the secondary component was apparent at phase 0.69. However, its D1 line was not apparent at phase 0.25 on 2020 Mar.~16; see Sect.~\ref{Sec:ChromActiv} for an explanation of the variability in the secondary component's sodium doublet spectra.

\begin{figure}
\includegraphics[width=\columnwidth]{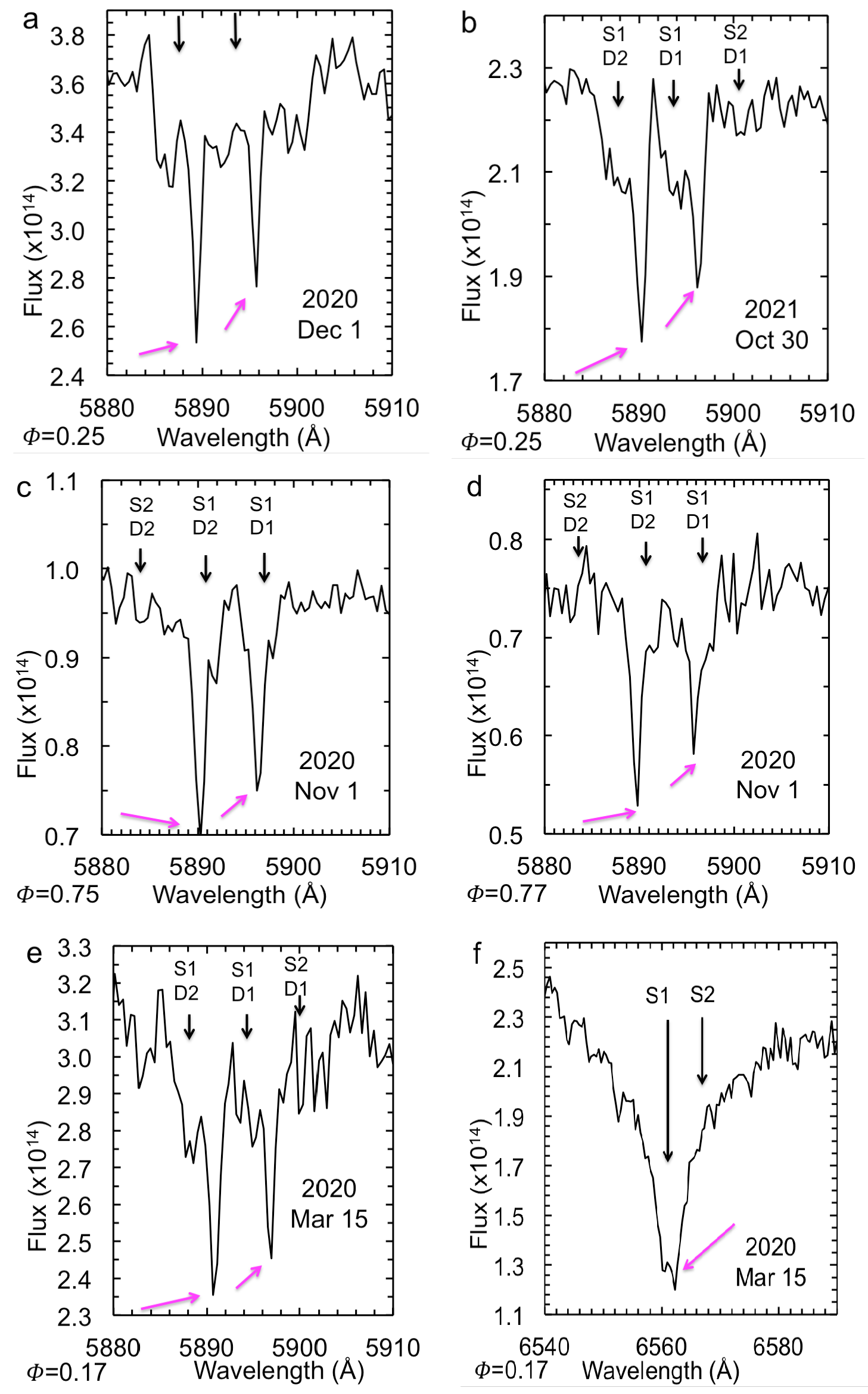}
\caption{Spectra of the Na~I~D doublet of RR~Lep during quadrature phases (a-e) and a corresponding H$_\alpha$ spectrum at phase 0.17~(f); $R=7000$. Exposure times: (a,~b)~240~s; (c)~360~s; (d)~120~s; (e, f)~240~s.}
\label{sodiumD_rrlep}
\end{figure}

With a maximum radial velocity of $-116\pm 10$~km~s$^{-1}$ (i.e. RV$_{\rm 1, total}={\rm \textit{K}}_1+V_0$; see Tables~\ref{Tab:Models_RRLep} and \ref{Tab:RVs}) at phase 0.25, the RR~Lep primary star’s sodium~D lines were distinct from those of the tertiary star (Fig.~\ref{sodiumD_rrlep}a,~b). In contrast, at phase 0.75 the positive radial velocity of the primary star was close to that of the tertiary star, resulting in blending of its spectral lines with those of the tertiary star. The lines of the primary star around phase 0.75 were broadened to the red of the sharp lines of the tertiary star (Fig.~\ref{sodiumD_rrlep}c,~d). The Na~I~D lines of the tertiary component were strong relative to those of the primary component at phase 0.17 (Fig.~\ref{sodiumD_rrlep}e). The tertiary H$_\alpha$ spectrum at phase 0.17 was blended with that of the primary component, although noticeable with broadening to the red (Fig.~\ref{sodiumD_rrlep}f). The secondary component’s flux was too weak for its H$_\alpha$ line to be apparent.

\subsection{Chromospherical activity}
\label{Sec:ChromActiv}

\begin{figure}
\centering
\includegraphics[width=8.4cm]{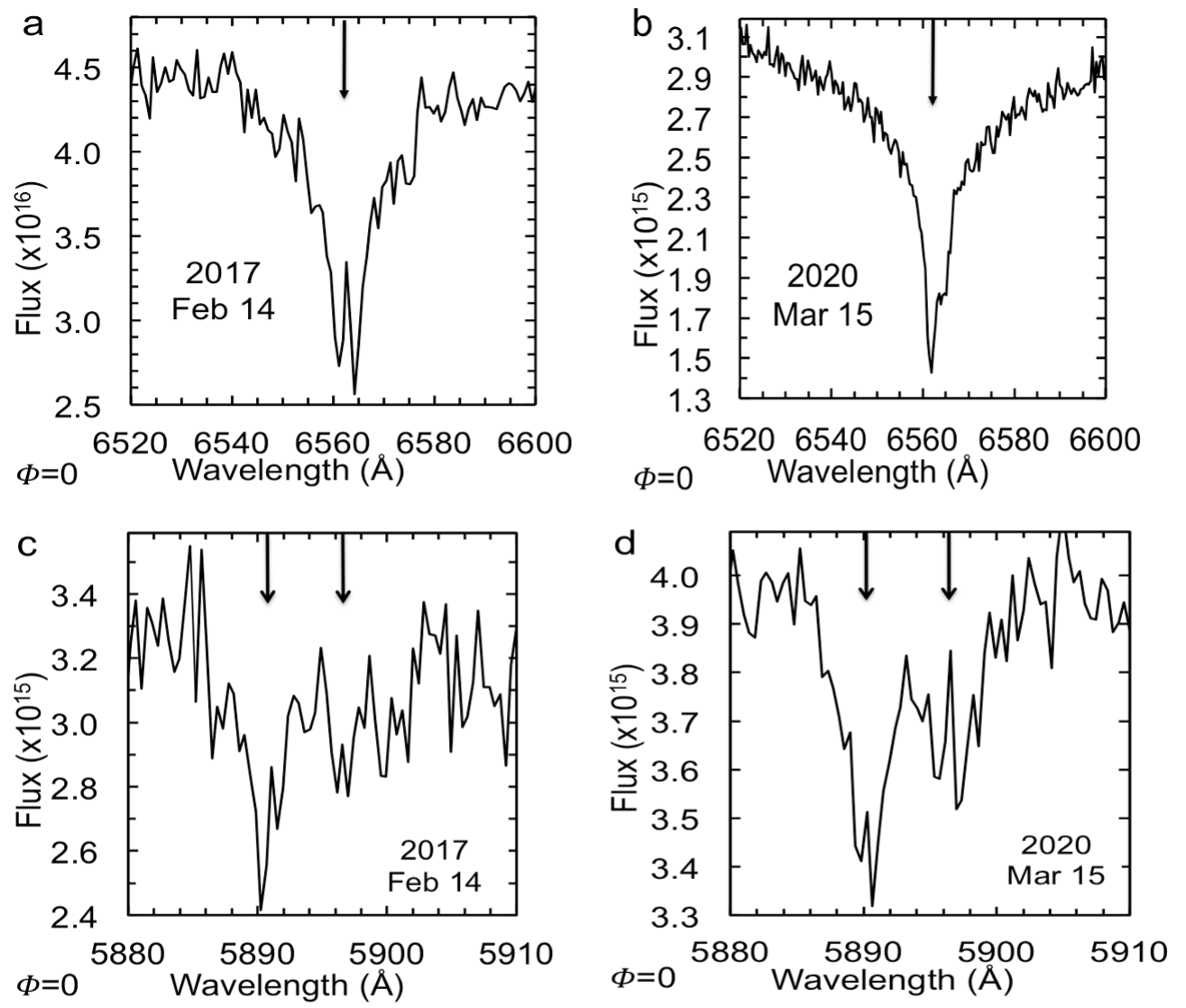}
\caption{Spectra of BF~Vel during primary eclipses. Indications of chromospherical activity in the secondary component of BF~Vel during primary eclipse with emission in the H$_\alpha$ line centre (a,~b) and emission in the Na~I~D lines on the same dates (c,~d). Exposure times were 60~s.}
\label{halpha_bfvel}
\vspace{2mm}
\includegraphics[width=8.4cm]{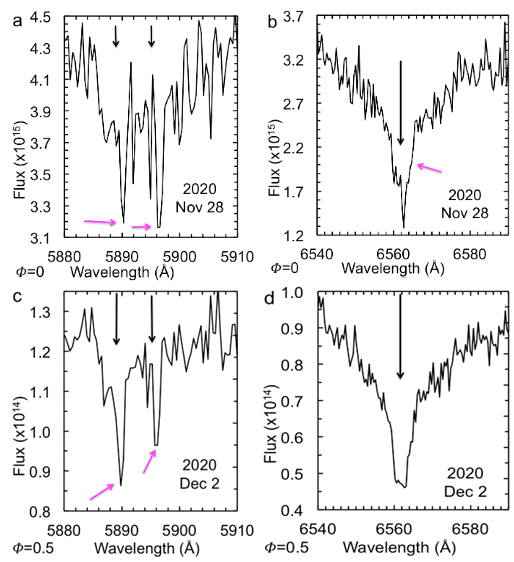}
\caption{Spectra of RR~Lep during eclipses. Downward arrows mark the calculated position of the combined binary components' lines. (a)~The magenta arrows indicate the line centres of the tertiary star's Na~I~D doublet. (b)~The H$_\alpha$ line is partly infilled broadened to the red; it is marked with a magenta arrow. (c)~Secondary eclipse; the Na~I~D lines of the primary star broadened to the red by the tertiary star (magenta). (d)~The H$_\alpha$ line broadened to the red. Exposure times were 60~s.}
\label{halpha_rrlep}
\end{figure}

The spectrum of BF~Vel during a primary eclipse is dominated by the secondary component, which has almost the same diameter as the primary star. With an orbital inclination of close to $90\degr$, the eclipse is annular. Emission in the H$_\alpha$ line centre of BF~Vel was observed during the primary eclipse on 2017 Feb.~14 (Fig.~\ref{halpha_bfvel}a); the H$_\alpha$ line centre was infilled on 2020 Mar.~15 (Fig.~\ref{halpha_bfvel}b). Emission was also observed on the same dates in the Na~I~D line centres of BF~Vel (Fig.~\ref{halpha_bfvel}c,~d). These spectra indicate the secondary component is chromospherically active. Evidence of chromospherical activity in the secondary component of BF~Vel was also apparent in the variability in its sodium~D spectral lines at other phases (Fig.~\ref{sodiumD_bfvel}). Its Balmer lines are weak compared to those of the primary star, whereas its sodium~D lines are stronger. The spectra also include blending of the tertiary component’s lines.

In contrast to BF~Vel, the spectral lines of the primary star of RR~Lep contribute to the spectrum during a primary eclipse as the orbital inclination is $\sim81\degr$. Spectra in the range 5700-–6700~{\AA} were reduced with a wavelength interval of 0.42~{\AA} to enhance resolution in the Na~I~D and H$_\alpha$ regions. The Na~I~D lines of the tertiary component in RR~Lep are strong and distinct from those of the binary components (Fig.~\ref{halpha_rrlep}a,~c). The infilling in the H$_\alpha$ line of RR~Lep at phases 0 and 0.5 indicates that both the primary and secondary components are chromospherically active (Fig.~\ref{halpha_rrlep}b,~d). The strength of these lines indicates that the spectral type of the tertiary star is about G8.


\section{Photometry}
\label{Sec:Photometry}

\subsection{Ground-based photometry}
\label{Sec:G-B_Photometry}

The system RR~Lep was observed with the 40~cm Cassegrain telescope of the University of Athens Observatory, Greece which is equipped with the ST-10XME CCD. The observations were made in the $B$ and $V$ pass bands (Bessell specification) in nine nights between January and March 2012. The light curves (LCs) have been already published in \citet{LIA13} but we included them in the present work as additional information for the modelling.

The Evans 0.51~m telescope at El~Sauce Observatory in Chile was used to observe BF~Vel in $BVRI$ Johnson-Cousins filters on 26 nights, commencing 13~Dec. 2022. The telescope was equipped with a Moravian C3-26000 CMOS camera. The first five data sets were observed at 2-min cadence in alternating $B$ and $V$ filters. Subsequent observations for $B$, $V$, and $R_{\rm c}$ filters were at 1-min cadence and $I_{\rm c}$ filter observations at 90-sec cadence. Standard calibration differential photometry was then performed on each data set using `AstroImageJ~V5' \citep{COL17} resulting complete LCs for $B$, $V$, and $R_{\rm c}$ filters and an almost complete LC for the $I_{\rm c}$ filter.

\subsection{TESS photometry}
\label{Sec:TESS_Photometry}

RR~Lep (TIC~169532543) was observed with a 2-min cadence in Sector~5 (Nov.~15 to Dec.~11, 2018) and with a 10-min cadence in Sector~32 (Nov.~20 to Dec.~16, 2020). BF~Vel (TIC~190894967) was observed with a 30-min cadence in Sector~8 (Feb.~2 to 27, 2019),
with a 30-min cadence in Sector~9 (Feb.~28 to Mar.~25, 2019), with a 2-min cadence in Sector~35 (Feb.~9 to Mar.~6, 2021), and with a 200-s cadence in Sector~62 (Feb.~12 to Mar.~10, 2023).
These TESS data were downloaded from the Mikulski Archive for Space Telescopes (MAST) \citep[cf.][]{JEN16}. The straight Simple Aperture Photometry (SAP) data were used, since the Pre-search Data Conditioning Simple Aperture Photometry (PDCSAP) is optimised for planet transits, and it appears that the PDCSAP detrending sometimes leads to additional effects. The LCs from Sector~5 for RR~Lep and Sector~35 for BF~Vel were used to ensure that a reasonably high Nyquist frequency could be sustained for the pulsation analysis. The Sector~35 includes the data in two orbits, as in other sectors. When comparing the data taken in these two orbits, the primary minimum depths are slightly different. This affects the residual LCs (and therefore the pulsation analysis) when the entire data set is resolved. Therefore, these two data obtained in two orbits were solved separately. Photometry of these two binary systems is publicly accessible \citep{RIC15}.

\section{Modelling radial velocity and light curves}
\label{Sec:Models}
The numerical integration method of \citet[][WD]{WIL71} was applied to solve the LCs and RV curves of BF~Vel and RR~Lep. This method models the RV curve and (or) the LC of a given binary star, considering mainly the ellipticity and proximity effects, and the equipotential surfaces of the components. Radial velocities were not determined with {\sc Ravespan} close to the orbital phases 0.0 and 0.5, where the components occult each other and proximity effects are dominant.

\begin{table*}
\begin{center}
\caption{Results of optimal curve-fitting to the radial velocity and light curves of BF~Vel using WD modelling.
\label{Tab:Models_BFVel}}
\scalebox{0.94}{
\begin{tabular}{lcccccc}
\hline\hline
Parameter		& 	$B$ & $V$ & $I$ & Sector 35-1    & Sector 35-2		& Sector 62 \\
\hline
$T_0$ 	& \multicolumn{3}{c}{2459266.8845 HJD}   &   \multicolumn{2}{c}{2459266.8853 BJD} \\
$P$ (d)		& \multicolumn{5}{c}{0.7040299}	 \\
\hline
$a$ (R$_{\sun}$) 	&	 4.76$\pm0.03$	&	 4.74$\pm0.05$	&	 4.77$\pm0.05$	&	 4.74$\pm0.04$	&	 4.74$\pm0.04$	&	 4.74$\pm0.04$ \\
$V_0$ (km s$^{-1}$)	&	 $4.1\pm1.1$	&	 5.3$\pm1.8$	&	 4.1$\pm1.2$	&	 5.1$\pm1.4$	&	 5.4$\pm1.4$	&	 5.4$\pm1.3$ \\				\\
$K_1$ (km s$^{-1}$)	&	116.4$\pm11.5$    & 113.0$\pm3.8$ & 116.5$\pm4.0$ & 113.0$\pm4.7$ 	& 113.0$\pm4.7$    & 113.0$\pm3.6$									\\
$K_2$ (km s$^{-1}$)	&	223.4$\pm17.8$    & 226.0$\pm3.1$ & 224.1$\pm3.4$ &  225.9$\pm4.8$    & 225.9$\pm4.9$ & 225.9$\pm2.7$										\\
$\Delta \phi$	&	 $-0.0003\pm0.0004$	&	 $-0.0001\pm0.0001$	&	 $-0.0002\pm0.0001$	&	 0.0002$\pm0.0001$	&	 0.0001$\pm0.0001$	&	 0.0010$\pm0.0001$   \\
$i$ ($\degr$)	&	 84.3$\pm0.8$	&	 84.5$\pm0.1$	&	 83.6$\pm0.1$	&	 84.4$\pm0.2$	&	 84.3$\pm0.2$	&	 84.3$\pm0.1$ \\
$T_1$ (K)	&	\multicolumn{6}{c}{8100 (fixed)}										\\
$T_2$ (K)	&	 4436$\pm116$	&	 4518$\pm24$	&	 4587$\pm20$	&	 4613$\pm19$	&	 4613$\pm18$	&	 4552$\pm17$  \\
$\Omega_1$	&	 3.55$\pm0.06$	&	 3.45$\pm0.02$	&	 3.49$\pm0.02$	&	 3.43$\pm0.02$	&	 3.43$\pm0.02$	&	 3.41$\pm0.01$ \\
$\Omega_2$	&	2.91	&	2.87	&	2.92	&	2.88	&	2.87	&	 2.87  \\
$q=M_2/M_1$ 	&	 0.52$\pm0.01$ 	&	 0.50$\pm0.01$	&	 0.52$\pm0.01$	&	 0.50$\pm0.01$	&	 0.50$\pm0.01$	&	 0.50$\pm0.01$ \\
$r_1$ (mean)	&	 0.34$\pm0.04$	&	 0.35$\pm0.02$	&	 0.34$\pm0.02$	&	 0.35$\pm0.02$	&	 0.35$\pm0.02$	&	 0.35$\pm0.01$  \\
$r_2$ (mean)	&	 0.32$\pm0.01$	&	 0.32$\pm0.01$	&	 0.32$\pm0.01$	&	 0.32$\pm0.01$	&	 0.32$\pm0.01$	&	 0.32$\pm0.01$  \\
$L_1$ 	&	 0.86$\pm0.02$	&	 0.86$\pm0.01$	&	 0.81$\pm0.01$	&	 0.74$\pm0.01$	&	 0.75$\pm0.01$  	&	 0.86$\pm0.02$  \\
$L_2$ 	&	 0.04$\pm0.01$	&	 0.07$\pm0.01$	&	 0.14$\pm0.01$	&	 0.13$\pm0.01$	&	 0.13$\pm0.01$ 	&	 0.14$\pm0.01$ \\
$L_3$ 	&	 0.10$\pm0.02$	&	 0.07$\pm0.01$	&	 0.05$\pm0.01$	&	 0.13$\pm0.01$	&	 0.12$\pm0.01$	&	 -- \\
\hline												
Spot parameters 	&		&		&		&		&		&	\\
$\beta$ (deg)	&	 -- 	&	 -- 	&	 -- 	&	 -- 	&	 -- 	&	 30$\pm7$      \\
$\lambda$ (deg)  	&	 -- 	&	 -- 	&	 -- 	&	 -- 	&	 -- 	&	 292$\pm4$      \\
$\gamma$ (deg)	&	 -- 	&	 -- 	&	 -- 	&	 -- 	&	 -- 	&	 30$\pm3$ \\
$\kappa$	&	 -- 	&	 -- 	&	 -- 	&	 -- 	&	 -- 	&	 0.85$\pm0.06$      \\
\hline												
$\chi^{2}_{\rm red}$ (RV$_1$)    	&	1.84	&	1.96	&	1.84	&	1.93	&	1.97	&	 1.96 \\
$\chi^{2}_{\rm red}$ (RV$_2$)    	&	1.32	&	1.31	&	1.33	&	1.31	&	1.31	&	 1.32 \\
$\nu$ (RV)       	&	20	&	20	&	20	&	20	&	20	&	 20  \\
$\Delta$RV (km~s$^{-1}$)  	&	12	&	12	&	12	&	12	&	12	&	 12     \\
\hline												
$\chi^{2}_{\rm red}$ (LC)    	&	1.96	&	1.07	&	1.04	&	1.6	&	1.51	&	 2.15  \\
$\nu$ (LC)       	&	1086	&	1246	&	1293	&	7379	&	5707	&	 9628  \\
$\Delta l$ (LC)  	&	0.004	&	0.004	&	0.004	&	0.002	&	0.002	&	 0.002 \\
\hline												
\end{tabular}}
\end{center}
\end{table*}

In the WD code (and most LC modelling programmes) the effective temperature ($T_1$) of the primary star is usually taken as a fixed parameter when solving the LCs of a given binary system. Therefore, reliable determination of $T_1$ is important. We examined three methods to determine the effective temperature of the primary components of BF~Vel and RR~Lep.\\
(i)~To calculate the intrinsic (unreddened) colours of both systems, the colour indices were first derived from the study of \citet{HOG00} to be $B-V=0.13\pm0.08$~mag for BF~Vel and $B-V=0.15\pm0.04$~mag for RR~Lep. Then, the colour excesses of $E(B-V)=0.095\pm0.015$~mag for BF~Vel and $E(B-V)=0.078\pm0.010$~mag for RR~Lep were computed using the stellar absorption values of $A_V=0.295\pm0.048$~mag for BF~Vel and $A_V=0.242\pm0.030$~mag for RR~Lep in the Gaia~DR3 catalogue \citep{GAIA22}. Finally, the intrinsic colours were calculated to be $(B-V)_0=0.035\pm0.081$~mag for BF~Vel and $(B-V)_0=0.072\pm0.041$~mag for RR~Lep.
When compared with the calibration data sets of \citet{PEC13} and \citet{EKE18}, these intrinsic colours indicate that the spectral types of these systems would be close to A1-2V, and therefore are not compatible with our spectral results.\\
(ii)~\citet{KAH24} reported the temperature of the primary component of RR~Lep as $7800\pm150$~K from the analysis of H$_\beta$ lines in the HERMES spectra, which they observed at phases close to the secondary eclipse.\\
(iii)~In Sect.~\ref{Sec:SpectrClass} of the present study, the spectral type of their primary components was assigned as A5V using the line matching method. From the calibration data sets of \citet{PEC13} and \citet{EKE18}, the temperature corresponding to the A5V spectral type is 8100~K and 8200~K, respectively. Therefore, the TESS LCs of BF~Vel were solved by using the temperatures of 8100~K and 8200~K for $T_1$, while the TESS LCs of RR~Lep were solved with temperatures of 7800~K, 8100~K, and 8200~K for $T_1$. Then, the temperature with the smallest chi-square value (i.e. 8100~K) was adopted and kept fixed for the primary components. The temperature of the secondary components was adjusted in the WD iterations.

\begin{table*}
\begin{center}
\caption{Results of optimal curve-fitting to the RV curves and light curves of RR~Lep using WD modelling.
\label{Tab:Models_RRLep}}
\begin{tabular}{lcccc}
\hline\hline
Parameter	&	$B$	&	$V$	&	\multicolumn{2}{c}{TESS} 	\\
            &       &       & Sector 5 & Sector 32 \\
\hline									
$T_0$ (HJD)	&	\multicolumn{2}{c}{2455953.2674}			&	\multicolumn{2}{c}{2459174.6486}			\\
$P$ (d)	&	\multicolumn{4}{c}{0.91542575}							\\
\hline									
$a$ (R$_{\sun}$)	&	 6.19$\pm0.03$	&	 6.11$\pm0.19$	&	 6.20$\pm0.03$	&	 6.17$\pm0.04$\\
$V_0$ (km~s$^{-1}$)	&	 $-43.5 \pm$0.7	&	 $-43.9 \pm$1.6	&	 $-44.0\pm0.8$	&	 $-44.0\pm1.0$\\
$K_1$ (km s$^{-1}$)	&	71.8$\pm9.4$  & 69.1$\pm6.1$  & 71.7$\pm3.7$  & 69.4$\pm4.9$						\\
$K_2$ (km s$^{-1}$)	&	265.8$\pm25.0$    & 265.6$\pm13.4$    & 265.6$\pm3.9$ & 266.8$\pm8.7$						\\
$\Delta\phi$	&	 $-0.0034 \pm0.0003$	&	 $-0.0020 \pm0.0002$	&	0.0007$\pm0.0001$	&	0.0002$\pm0.0001$\\
$i$~($\degr$)	&	 80.8$\pm$0.6	&	 82.4$\pm$0.4	&	79.9$\pm0.1$	&	80.4$\pm0.2$\\
$T_1$~(K)	&	\multicolumn{4}{c}{8100 (fixed)}						\\
$T_2$~(K)	&	 5005$\pm$51	&	 4870$\pm$38	&	4346$\pm18$	&	4467$\pm20$\\
$\Omega_1$	&	 2.82$\pm$0.02	&	 2.84$\pm$0.02	&	2.80$\pm0.07$	&	2.80$\pm0.01$\\
$\Omega_2$	&	2.41	&	2.37	&	2.39	&	2.39\\
$q=M_2/M_1$	&	 0.27$\pm$0.01	&	 0.26$\pm$0.01	&	0.27$\pm0.01$	&	0.26$\pm0.01$\\
$r_1$~(mean)	&	 0.40$\pm$0.02	&	 0.40$\pm$0.02	&	0.41$\pm0.01$	&	0.41$\pm0.01$\\
$r_2$~(mean)	&	 0.27$\pm$0.01	&	 0.27$\pm$0.02	&	0.27$\pm0.01$	&	0.27$\pm0.01$\\
$L_1$	&	 0.86$\pm$0.02	&	 0.82$\pm$0.02	&	0.89$\pm0.02$	&	0.88$\pm0.02$\\
$L_2$	&	 0.04$\pm$0.01	&	 0.06$\pm$0.01	&	0.07$\pm0.01$	&	0.07$\pm0.01$\\
$L_3$	&	 0.10$\pm$0.02	&	 0.12$\pm$0.02	&	0.04$\pm0.005$	&	0.05$\pm0.008$\\
\hline								
$\chi^{2}_{\rm red}$~(RV$_1$)	&	1.31	&	1.86	&	1.16	&	1.33\\
$\chi^{2}_{\rm red}$~(RV$_2$)	&	1.37	&	1.41	&	1.46	&	1.45\\
$\nu$~(RV)	&	17	&	17	&	17	&	17\\
$\Delta$RV$_1$~(km~s$^{-1}$)	&	4	&	4	&	4	&	4\\
$\Delta$RV$_2$~(km~s$^{-1}$)	&	12	&	12	&	12	&	12\\
\hline								
$\chi^{2}_{\rm red}$~(LC)	&	2.82	&	2.35	&	2.04	&	 1.42\\
$\nu$~(LC) 	&	1014	&	970	&	17230	&	 3496\\
$\Delta l$~(LC)	&	0.007	&	0.007	&	0.003	&	 0.003\\
\hline								
\end{tabular}
\end{center}
\end{table*}

Other parameters that were adjusted in the WD iterations included: the semi-major axis ($a$), mass ratio ($q$), and orbital inclination ($i$) of the binary system, phase shift ($\Delta \phi$; which allows the WD code to adjust for a zero point error in the ephemeris used to compute the phases; the unit is the orbital period), systemic velocity of the binary ($V_0$), non-dimensional surface potential parameter of each component ($\Omega _1$, $\Omega _2$), fractional luminosity of the primary component ($L_1$), and third light contribution to the total light of the system ($L_3$).

A quadratic limb-darkening law was assumed; the limb-darkening coefficients ($x_{\rm i}$ and $y_{\rm i}$) of the components were taken from \citet{CLA17} for the TESS LCs. For the ground-based LCs, the limb darkening coefficients were taken from the tables of \citet{CLA11}, according to the effective temperature of the components and the filters used. The bolometric gravity darkening exponents were adopted from \citet{LUC67} to be 0.32 for convective atmospheres of the components ($T \textless 7200$~K) and from \citet{ZEI24} to be 1.0 for radiative atmospheres of the components ($T \geq 7200$~K). The bolometric albedos of the components were set to 0.5 for convective atmospheres and to 1.0 for radiative atmospheres \citep{RUC69}. These parameters were kept constant during all iterations.

The iterations were started in MODE~2 in the WD code, which corresponds to a detached configuration; however, the final simultaneous RV+LC solution was switched to MODE~5, which applies to semi-detached binaries where the secondary component fills its Roche lobe. Therefore, the cited $\Omega _2$ retains its lobe-filling value and was not adjusted in this model.

\begin{figure}[h!]
\centering
\includegraphics[width=8.8cm]{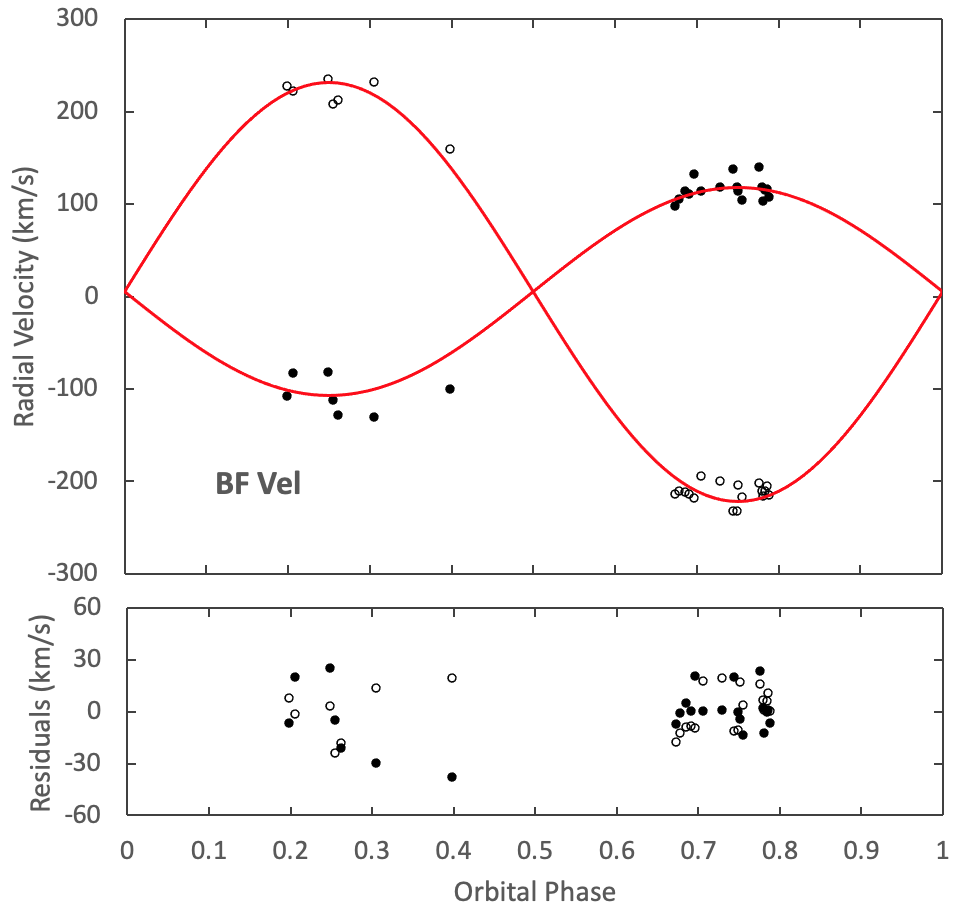}
\caption{Radial velocities of BF~Vel with the WD model fitting. RVs of the primary and the secondary components are marked as filled and hollow symbols, respectively. Residuals to the model are plotted in the bottom panel.
\label{fig:bfvel_rv}}
\includegraphics[width=8.7cm]{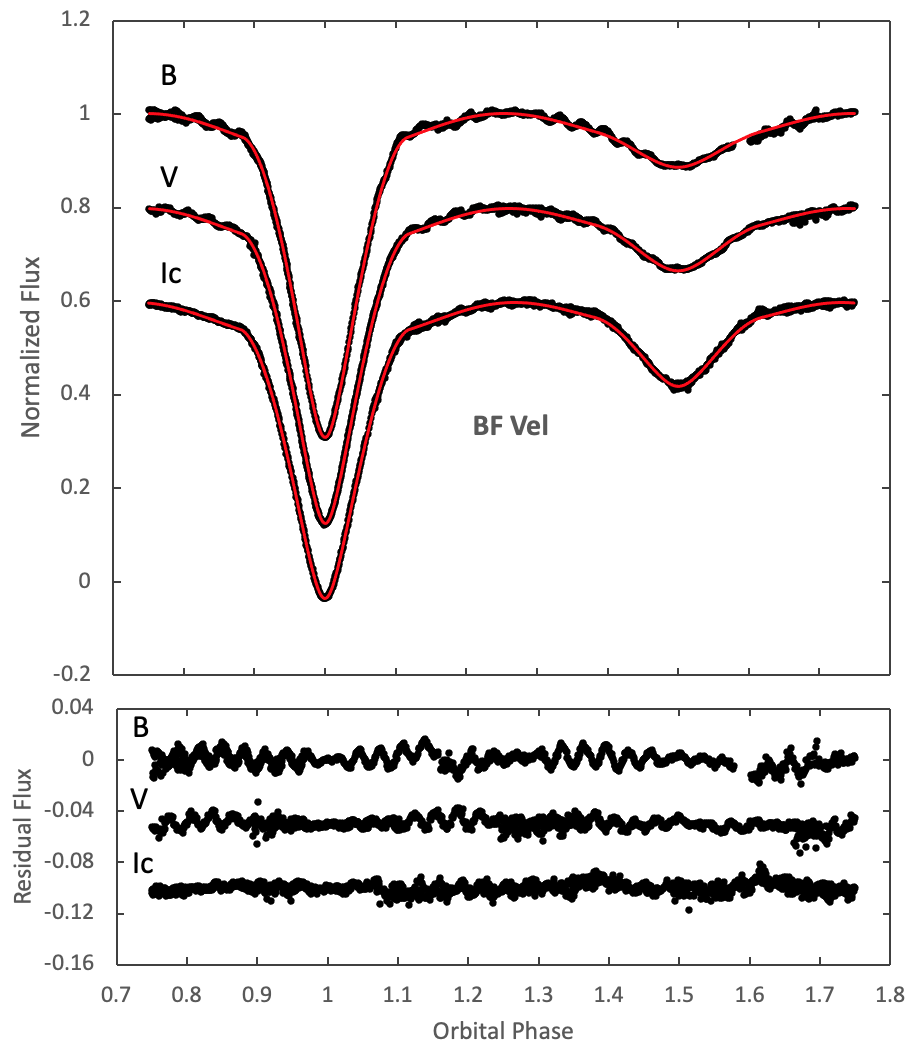}
\caption{Ground-based $BVI$ light curves of BF~Vel with the WD model fitting. Residuals to the model are plotted on the bottom panel. The $V$ and $I$ light curves and their residuals are shifted downward to enhance visibility.
\label{fig:bfvel_bvi_lc}}
\end{figure}

\begin{figure}
\centering
\includegraphics[width=8.2cm]{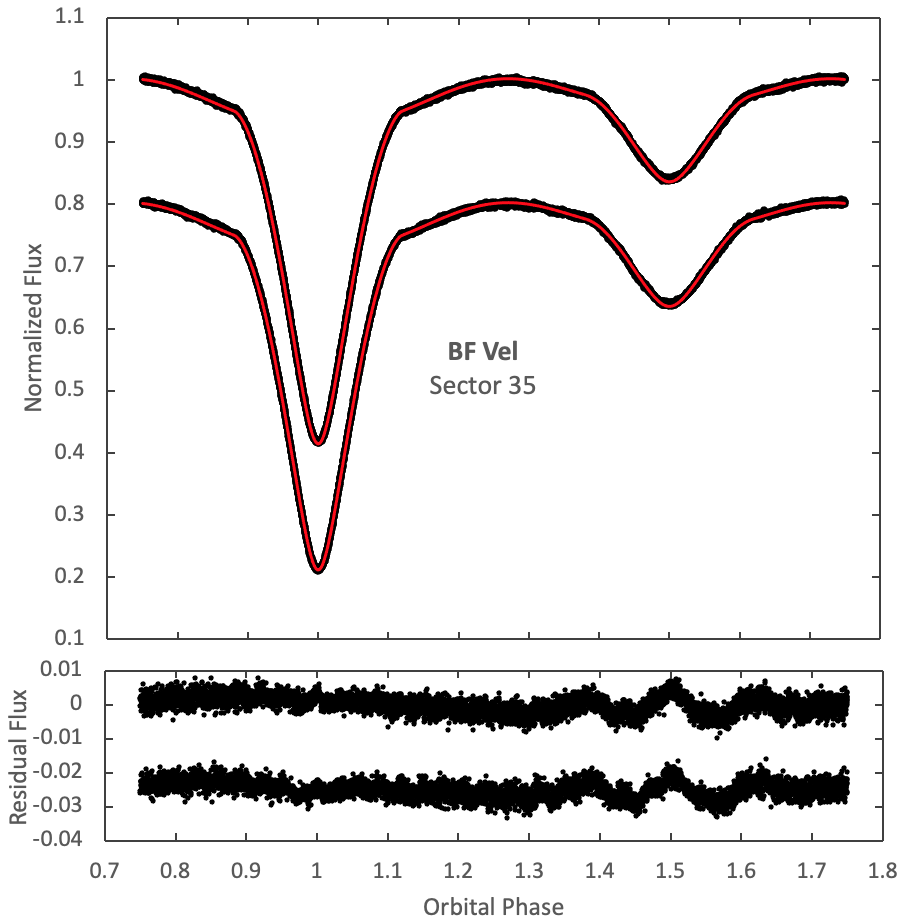}
\caption{Two light curves of TESS Sector~35 (with a 2-min cadence) obtained in two different consecutive orbits of BF~Vel and the WD model fitting. Residuals to the model are plotted on the bottom panel.
\label{fig:bfvel_tess_lc}}
\centering
\includegraphics[width=8.2cm]{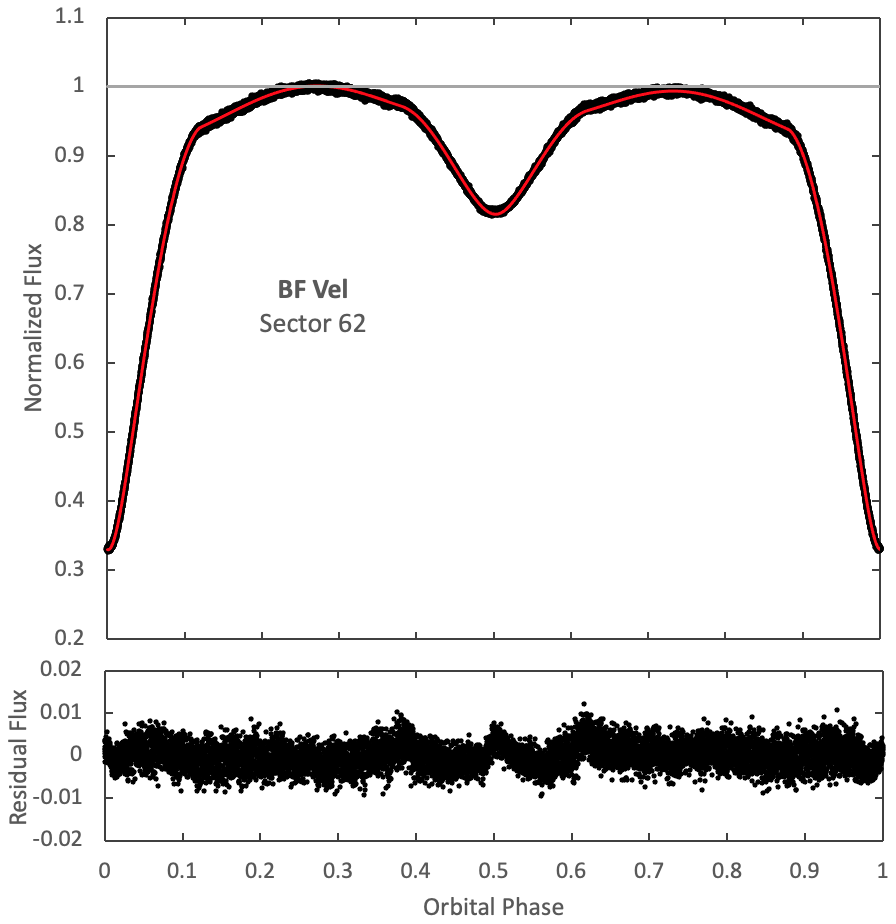}
\caption{Light curves of TESS Sector~62 (with a 200-s cadence) for BF~Vel and the WD model fitting. A solid gray line was added to the light flux value 1.0 to display the asymmetry between maxima in the light curve. Residuals to the model are plotted on the bottom panel.
\label{fig:bfvel_sec62_lc}}
\centering
\includegraphics[width=6cm]{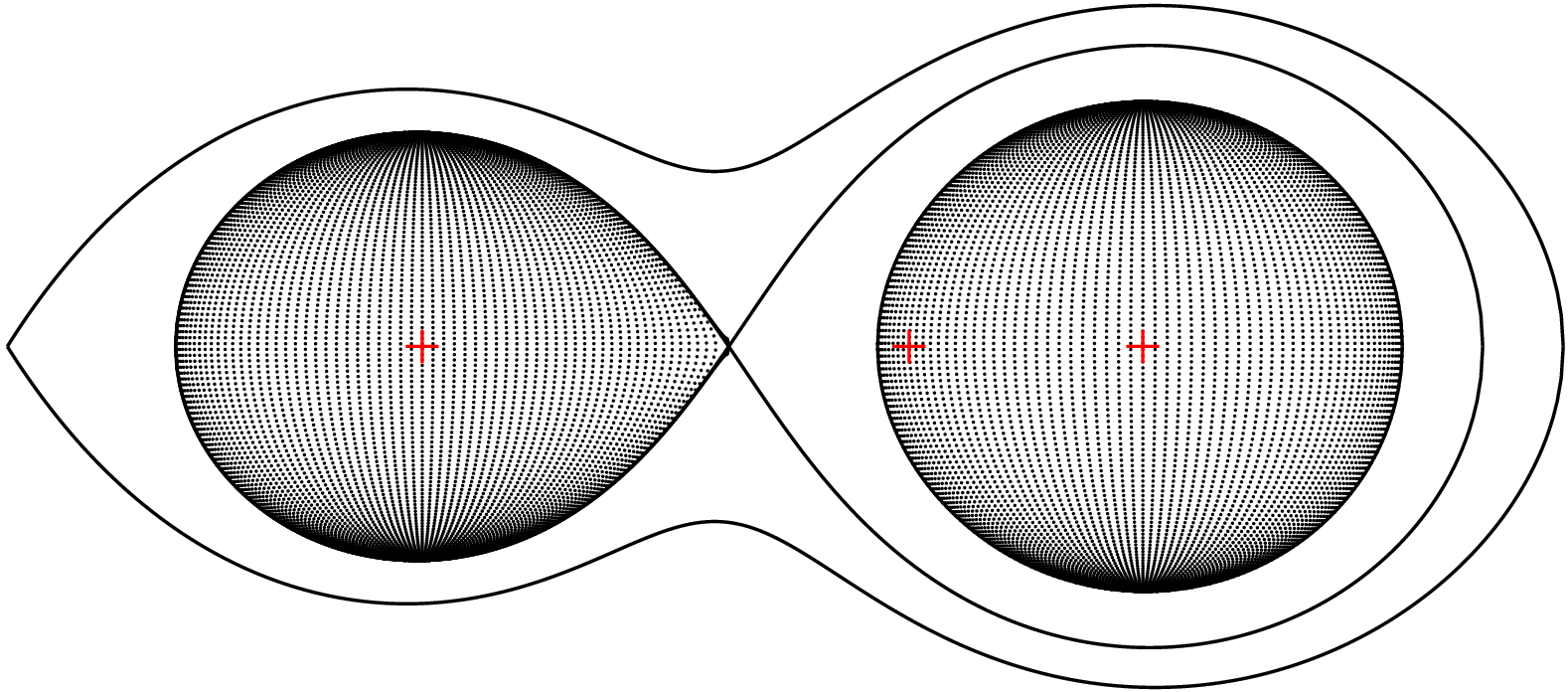}\\
\caption{Roche geometry of BF~Vel at orbital phase 0.75. Crosses denote the barycenters of the stars and the system.
\label{fig:bfvel_roche}}
\end{figure}

\begin{figure}
\centering
\includegraphics[width=8.7cm]{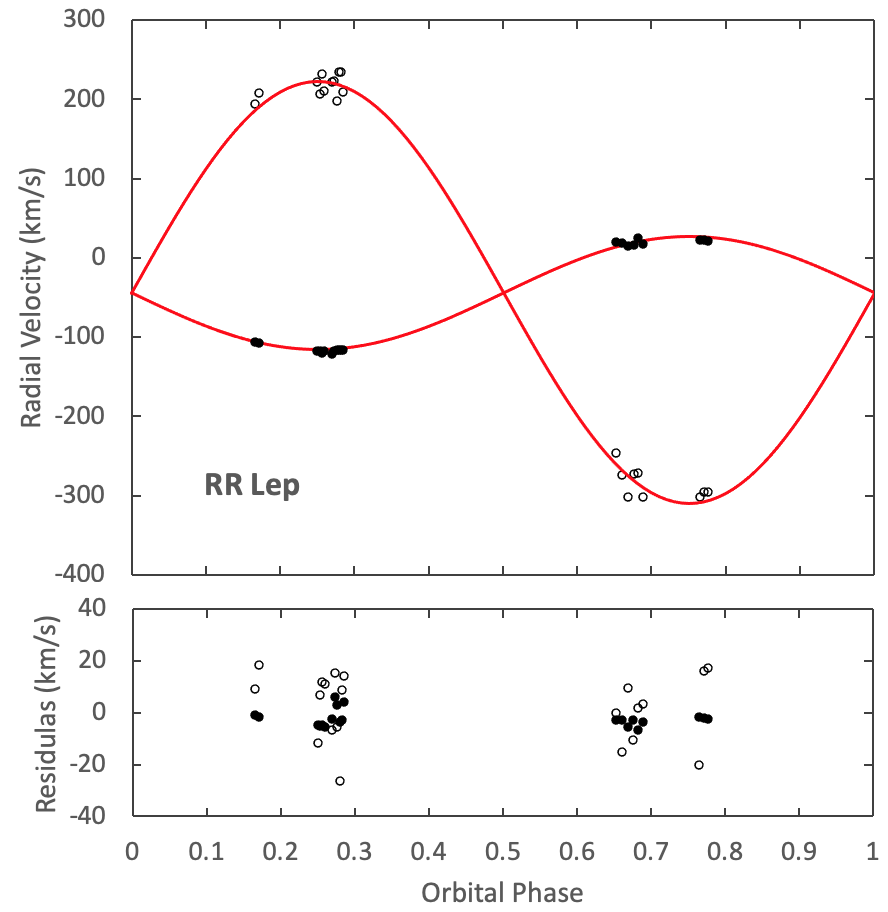}
\caption{Radial velocities of RR~Lep with the WD model fitting. RVs of the primary and the secondary components are marked as filled and hollow symbols, respectively. Residuals to the model are plotted in the bottom panel.
\label{fig:rrlep_rv}}
\includegraphics[width=8.7cm]{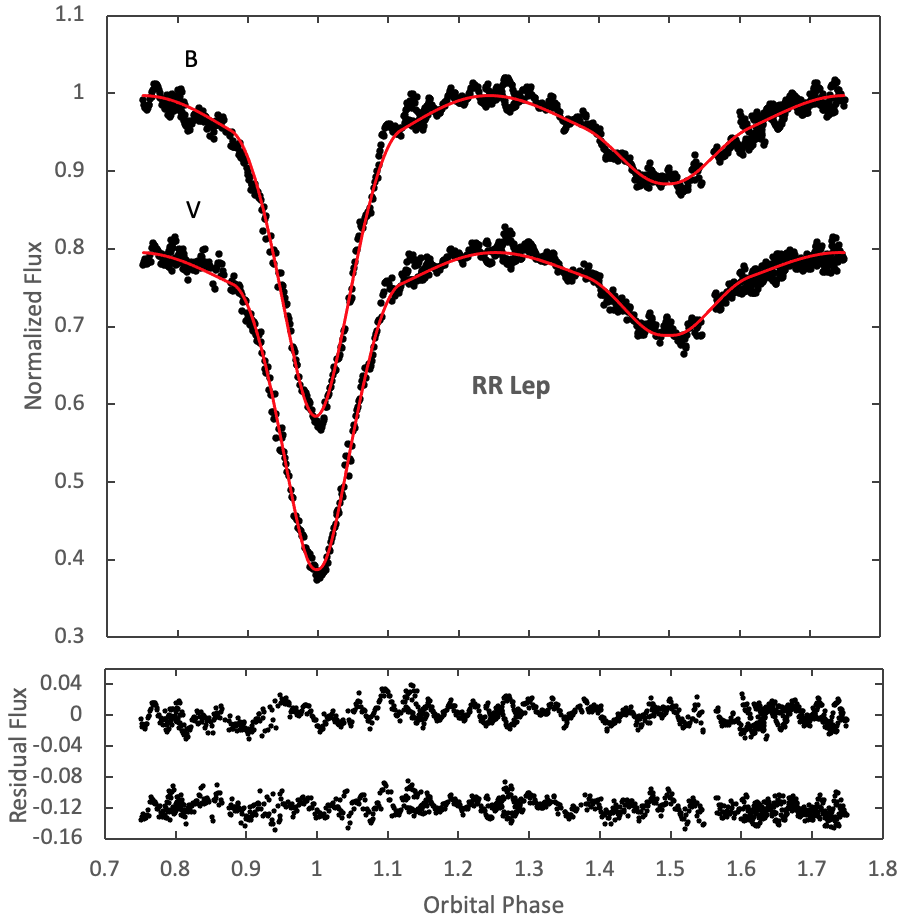}
\caption{Ground-based $BV$ light curves of RR~Lep with the WD model fitting. Residuals to the model are plotted on the bottom panel. The $V$ LC and its residuals have been shifted downward to enhance visibility.
\label{fig:rrlep_bv_lc}}
\end{figure}

\begin{figure}
\centering
\includegraphics[width=8.3cm]{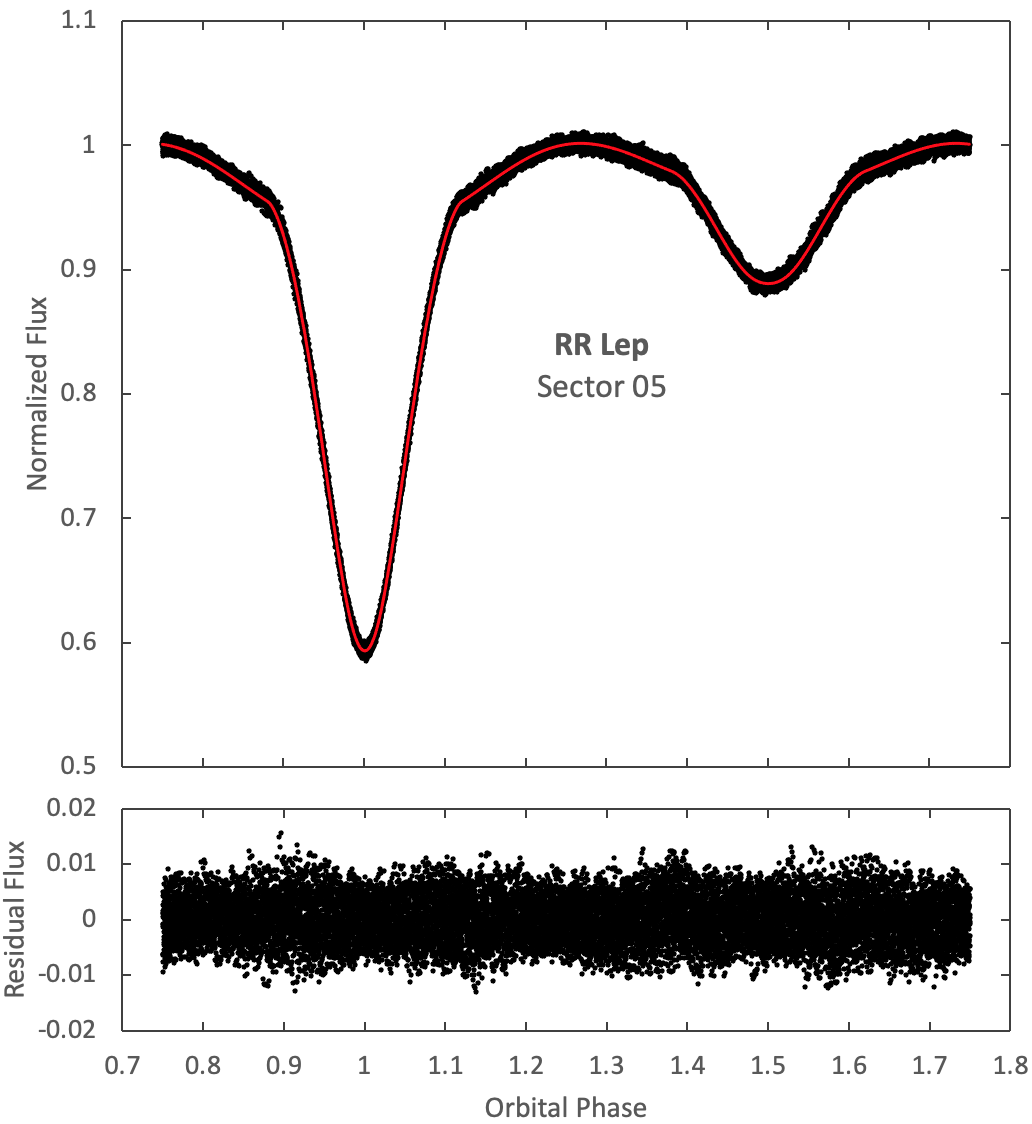}
\includegraphics[width=8.3cm]{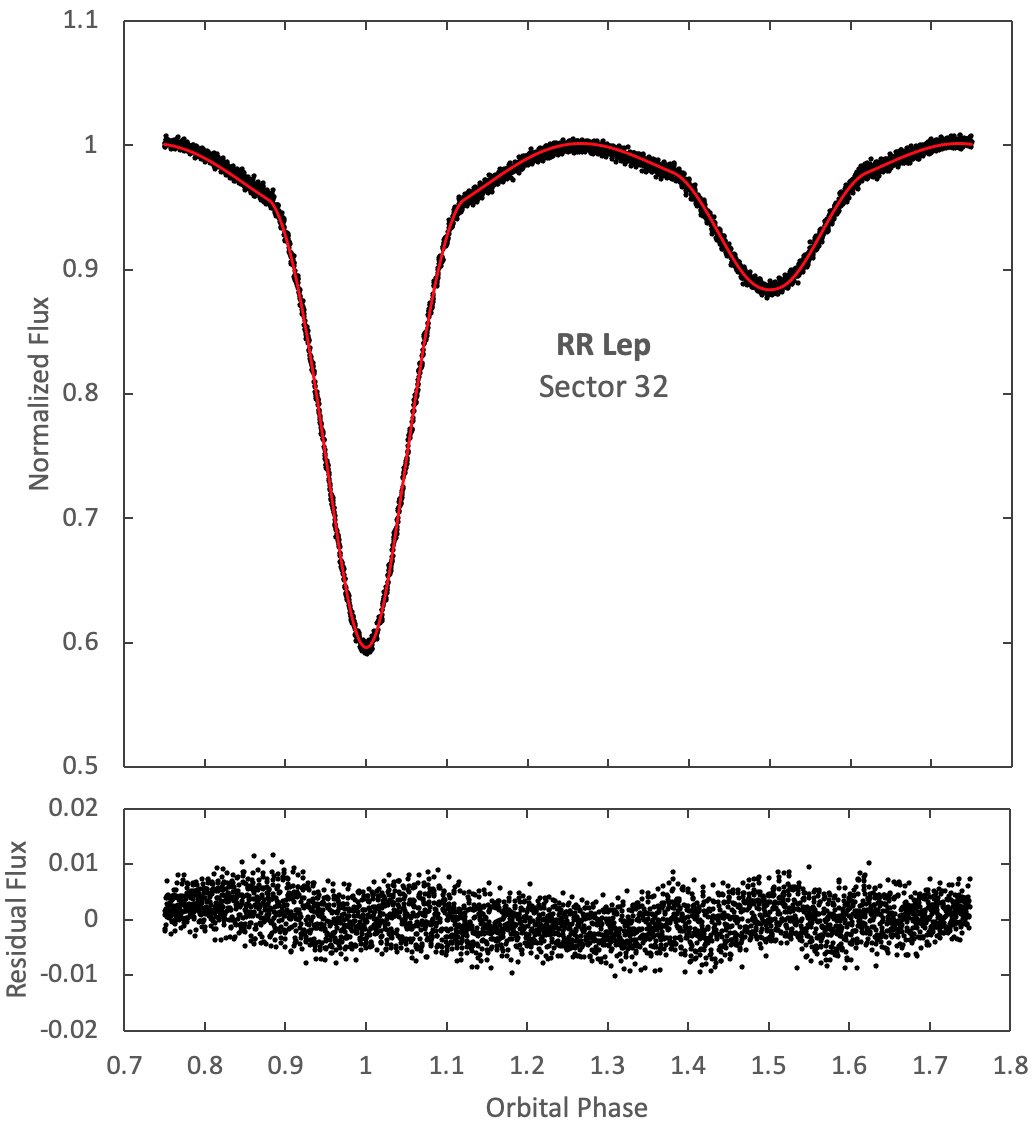}
\caption{TESS Sector~5 (with a 2-min cadence; upper panel) and Sector~32 (with a 10-min cadence; lower panel) light curves of RR~Lep with the WD model fitting. Residuals to the model are plotted in the bottom figures.
\label{fig:rrlep_lc}}
\end{figure}

\begin{figure}
\centering
\includegraphics[width=6.5cm]{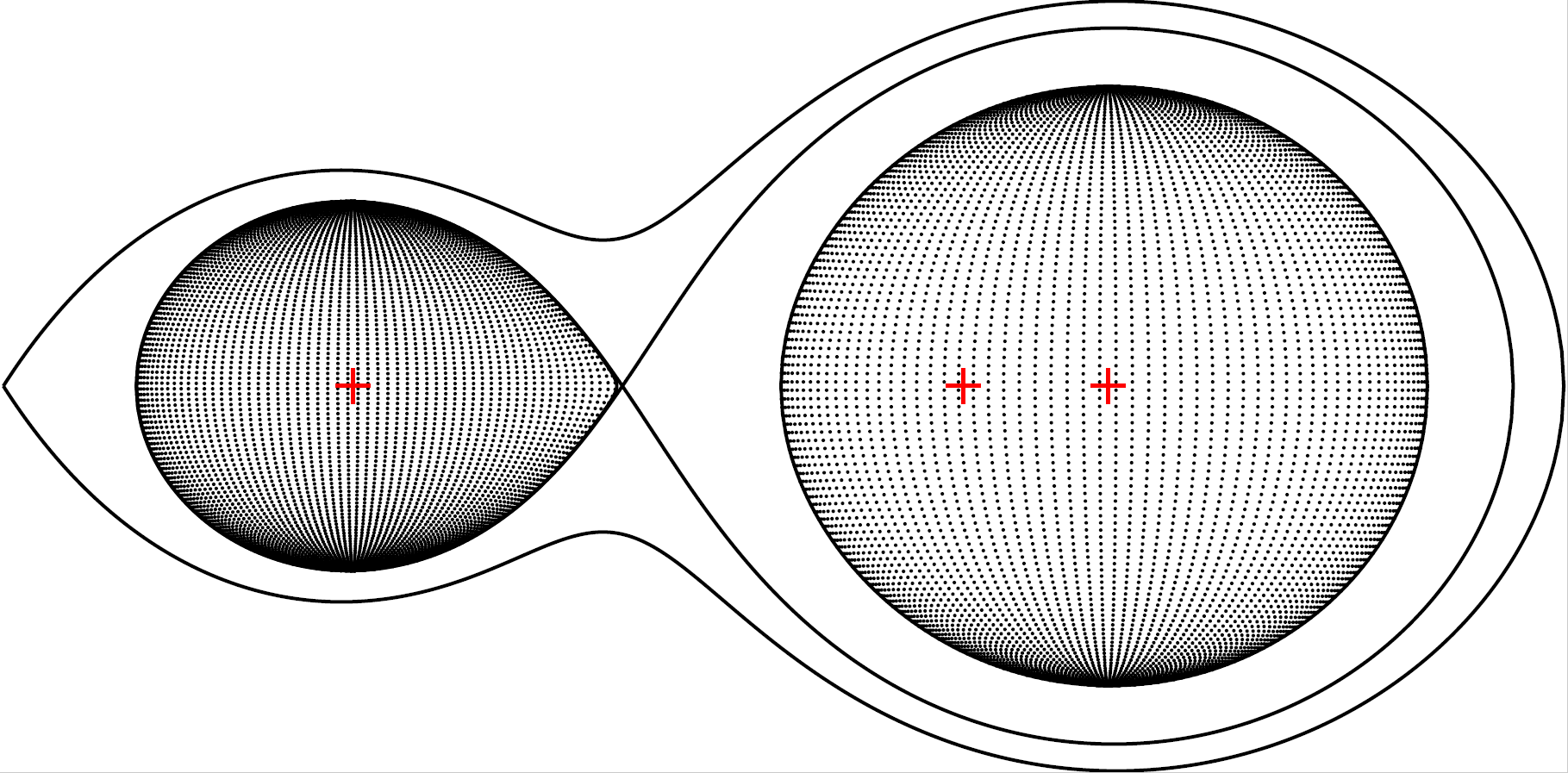}
\caption{Roche geometry of RR~Lep at orbital phase 0.75. Crosses denote the barycenters of the stars and the system.
\label{fig:rrlep_roche}}
\end{figure}

The parameters of the final WD models for the RVs and ground-based and TESS LCs of BF~Vel and RR~Lep are listed in Tables~\ref{Tab:Models_BFVel} and \ref{Tab:Models_RRLep}, respectively. The first two rows of these tables contain the ephemeris elements used for phasing both the LC and RV data. Since the amplitudes of the RVs, $K_1$ and $K_2$, are not directly given by the WD method, they were calculated using the relevant parameters in the RV+LCs fitting. The value of $\chi^{2}$ was obtained from $\sum_{\rm i}(\emph{y}_{\rm i,o}-\emph{y}_{\rm i,c})^{2}$/$\Delta\emph{y}_{\rm i}^{2}$ \citep{BEV69}, where  $y_{\rm i,o}$ and $y_{\rm i,c}$ are the observed and calculated light levels (or RVs) at a given phase, respectively, and $\Delta\emph{y}_{\rm i}$  is an error estimate for the measured values of $y_{\rm i,o}$. The reduced $\chi$-squared value was obtained from $\chi^{2}_{\rm red}$=$\chi^{2}/\nu$, where $\nu$ is the number of degrees of freedom of the data set, that is $\nu = N_{\rm obs} - N_{\rm param}$, where $N_{\rm obs}$ is the number of observations and $N_{\rm param}$ is the number of parameters in the model. The comparison of the RV measurements and photometric observations with the WD models are shown in Figs.~\ref{fig:bfvel_rv}, \ref{fig:bfvel_bvi_lc}, \ref{fig:bfvel_tess_lc}, and \ref{fig:bfvel_sec62_lc} for BF~Vel and in Figs.~\ref{fig:rrlep_rv}, \ref{fig:rrlep_bv_lc}, and \ref{fig:rrlep_lc} for RR~Lep. The Roche geometries of BF~Vel and RR~Lep, acquired using the program {\sc Binary Maker} \citep{BRA02}, are plotted in Figs.~\ref{fig:bfvel_roche} and \ref{fig:rrlep_roche}, respectively. In these configurations, which show that BF~Vel and RR~Lep are semi-detached binary systems, their primary components fill 84\% and 85\% of their Roche lobe, respectively, while the secondary components fill their Roche lobe completely.

In the WD method, the photometric data observed in various bands for a given binary star can be resolved either separately or all together simultaneously (with the RV data, if any). Due to the systemic uncertainties caused by different permeabilities of the filters, different pixel sensitivity and other instruments used, the model obtained from simultaneous solutions of various band photometric data cannot represent the observations better than the model derived from separate band solutions. Since we will perform a pulsation analysis using the residuals obtained after subtracting the eclipsing model from the observed LCs in this study, we solved each band's observed LC separately.

As can be seen in Table~\ref{Tab:Models_BFVel} for BF~Vel, there is no significant difference or discrepancy within the error limits between the model parameters in single-band LC+RV solutions. However for RR~Lep, values for the temperature of the secondary component ($T_{2}$) resulting from the solution of ground-based photometric observations are higher than those obtained from TESS solutions (according to Table~\ref{Tab:Models_RRLep}). In particular, in the solution of the ground-based $B$-band light curve, $T_{2}$ is $\sim 600$~K higher than temperatures given in the solution of TESS data. The CROWDSAP parameter, which is the ratio of the target flux to the total flux in the photometric aperture, is 0.985 for TESS Sector~5 and and 0.986 for Sector~32. These values show that only approximately 1.5\% of the observed flux in the SAP aperture does not come from the target (and the analysis would just add that to the third light), so this does not seem to be the reason. On the other hand, when the third light was neglected in the solution of the ground-based $B$ light curve, $T_{2}$ decreases to 4900~K, but, unfortunately, the difference was still present. When all light and RV curves were solved simultaneously, a value of 4650~K was obtained for $T_{2}$. However, the theoretical curve of this solution did not match the $B$-band observations particularly well (i.e. the fitting was not good). Therefore it seems that, the higher temperature of the secondary component in the ground-based photometric model may be due to systematic errors in the ground-based data.

In the TESS LC of Sector~62 of BF Vel there is a noticeable asymmetry at its maximum values (see Fig.~\ref{fig:bfvel_sec62_lc}). The light flux at Maximum~II ($\Phi=0.75$) is slightly lower than at Maximum~I ($\Phi=0.25$). Two different approaches were used to model this asymmetry: (i) The assumption of a cool spot on the secondary component due to the chromospherical activity detected in Sect.~\ref{Sec:ChromActiv}. (ii) The assumption of a hot spot that may form on the primary component due to mass transfer from the secondary to the primary, which causes the orbital period of the system BF~Vel to increase in the modelled parabolic change in the ETV analysis in Sect.~\ref{Sec:ETV_BFVel} and fits the semi-detached configuration. The WD code combined with the Monte Carlo (MC) optimization procedure \citep{ZOL04} were used to solve the TESS Sector~62 LC under these two different spot approaches. The most important contribution of the MC procedure in the WD+MC method is the use of a range of input values for the free parameters, rather than fixed input values. Thus, the WD+MC method searches for the model that provides the best fit of the observed LC in the solution space, by conducting hundreds of thousands of iterations in the selected input ranges. A Chi-square test was also performed using the model and observed data, and it was found that the cool spot model reflects the observations better. The results are given in Table~\ref{Tab:Models_BFVel}, and a comparison of the observed LC and the cool spot model synthetic LC is presented in Fig.~\ref{fig:bfvel_sec62_lc}. The spot parameters given in Table~\ref{Tab:Models_BFVel} are: The spot's latitude~($\beta$), longitude~($\lambda$), angular radius~($\gamma$), and temperature factor~($\kappa$). As a LC solution only was developed with WD+MC for the Sector 62 data, there are no RV curve solution parameters in the last column of Table~\ref{Tab:Models_BFVel}. When the LCs of TESS Sectors~35 and 62 are compared with each other, it is apparent that the depths of minima are quite different. The depth of Minimum~I ($\Phi=0.0$) was determined as 0.95~mag for Sector~35 and 1.20~mag for Sector~62, while the depth of Minimum~II ($\Phi=0.5$) was determined as 0.76~mag for Sector~35 and 0.98~mag for Sector~62. The reason for the different eclipse depths in the TESS data might be caused by contamination within the photometric aperture. The CROWDSAP parameter for the Sector~35 is 0.916, which indicates that is 8.4\% of the flux in the aperture does not come from the target. Unfortunately, this parameter is not available for the Sector~62 data. This difference in the WD+MC iterations is attributed to the third light ($L_3$) and not to the spot effect for Sector~62 data, although the third light contribution is at the level of 0.11 for Sector~35 data, it reduced $L_3$ to zero for Sector~62 data. On the other hand, while the fractional luminosity of the primary component ($L_1$) was 0.76 for Sector~35 data, $L_1$ increased to 0.87 for Sector~62 data. Thus, the third light contribution of 0.11, which cannot be seen in Sector~62, was included with $L_1$ in WD+MC results. Apart from this, the fact that the other parameters representing the LC for the Sector 62 data were similar to the parameters for Sector~35 data and other band data within error limits, does not change the results of the general solution for BF~Vel.


\section{Absolute Parameters and Evolutionary status}
\label{Sec:AbsPar}

The physical parameters of the components of both systems were calculated based on simultaneous modelling of the $V$-band LC and RVs (see Sect.~\ref{Sec:Models}) in order to simplify calculation of the bolometric magnitude ($M_{\rm bol}=M_{\rm V}+BC_{\rm V}$), and then derive the total luminosity of the system ($M_{\rm bol}=M_{\rm bol,\sun}-2.5\log L/L_{\sun}$) and the distance to the system from the distance module ($M_{\rm V}=V+5\log \pi +5 -A_{\rm V}$). The physical parameters thus obtained are given in Table~\ref{table:ABSPAR} along with their errors. The nominal solar values, which were adopted by IAU 2015 Resolutions B2 and B3, were used in our calculations. Bolometric corrections for the components were taken from the study of \citet{EKE18}, according to their effective temperatures.

\begin{table}
\centering
\caption{Absolute parameters of RR~Lep and BF~Vel.}
\label{table:ABSPAR}
\begin{tabular}{lcc}
\hline
\hline
Parameter		& RR Lep		& BF Vel	\\
\hline
$a$~(R$_{\sun}$)	&	 $6.11\pm0.19$	&	 $4.74 \pm0.05$  \\		
$M_1$~(M$_{\sun}$) 	&	 $2.90 \pm0.36$	&	 $1.93 \pm0.09$ \\		
$M_2$~(M$_{\sun}$) 	&	 $0.75 \pm0.12$	&	 $0.97 \pm0.06$ \\		
$R_1$~(R$_{\sun}$) 	&	 $2.44 \pm0.20$	&	 $1.66 \pm0.11$ \\		
$R_2$~(R$_{\sun}$)	&	 $1.65 \pm0.17$	&	 $1.52 \pm0.06$ \\		
$\log g_1$~(cm~s$^{-2}$)	&	 $4.13 \pm0.08$	&	 $4.28 \pm0.06$ \\		
$\log g_2$~(cm~s$^{-2}$) 	&	 $3.88 \pm0.06$	&	 $4.06 \pm0.02$ \\		
$T_1$ (K) 	&	 $8100 \pm300$	&	 $8100 \pm300$ \\		
$T_2$ (K) 	&	 $4870 \pm120$	&	 $4518 \pm115$ \\		
$L_1$~(L$_{\sun}$)	&	 $23.1 \pm4.3$	&	 $10.7 \pm1.4$ \\		
$L_2$~(L$_{\sun}$)  	&	 $1.38 \pm0.22$	&	 $0.87 \pm0.11$ \\		
$M_{\rm bol,1}$ (mag) 	&	 $1.33 \pm0.34$	&	 $2.17 \pm0.31$ \\		
$M_{\rm bol,2}$ (mag) 	&	 $4.39 \pm0.33$	&	 $4.90 \pm0.20$ \\		
$M_{\rm V,1}$ (mag) 	&	 $1.32 \pm0.34$	&	 $2.16 \pm0.31$ \\		
$M_{\rm V,2}$ (mag) 	&	 $4.71 \pm0.33$	&	 $5.44 \pm0.20$ \\		
$A_{\rm V}^{a}$ (mag) 	&	 $0.242 \pm0.030$	&	 $0.295 \pm0.048$ \\		
$E$($B-V$) (mag) 	&	 $0.078 \pm 0.010$   	&	 $0.095 \pm0.015$		 \\
$B-V^{b}$ (mag) 	&	 0.15 $\pm$0.04  	&	 0.13 $\pm$ 0.08		\\
$V^{b}$ (mag) 	&	 10.14 $\pm$ 0.03    	&	 10.93 $\pm$ 0.06 		\\
$M_{\rm V, system}$ (mag) 	&	 $1.25\pm$0.32 	&	 $2.08\pm$ 0.29	\\	
$d$ (pc) 	&	 537 $\pm$ 42  	&	 514 $\pm$ 32	\\	
\hline
\end{tabular}
\\$^{a}$\citet{CRE23}; $^{b}$\citet{FAB02}
\end{table}

The results of the RV+LC analyses show that the total luminosity of RR~Lep is 28.17~L$_{\sun}$ and that of BF~Vel is 12.43~L$_{\sun}$ (see Tables~\ref{Tab:Models_BFVel}, \ref{Tab:Models_RRLep}, and \ref{table:ABSPAR}). In the Gaia DR3 catalogue, the  system luminosities produced from the FLAME module are given as 37.32~L$_{\sun}$ for RR~Lep and 12.74~L$_{\sun}$ for BF~Vel.

An absolute magnitude of $M_{\rm G}=1.97$~mag for BF~Vel was calculated using the $G$-band mean magnitude $m_{\rm G}=10.77$~mag, parallax $\pi=1.9776$~mas from the Gaia~DR3 catalogue and the interstellar absorption (or extinction) $A_{\rm G}=0.275$~mag, produced from the ESP-HS module \citep{CRE23}. Based on these data, the BF~Vel system luminosity is 12.78~L$_{\sun}$, assuming zero bolometric correction. The system's luminosity and the distance value calculated from the $V$-band LC+RV solution for BF~Vel (12.43~L$_{\sun}$ and $514\pm32$~pc) agree with the values given in the Gaia~DR3 catalog (12.74 or 12.78~L$_{\sun}$ and $506\pm7$~pc).

In a similar manner for RR~Lep, using $m_{\rm G}=9.94$~mag, $\pi=2.4453$~mas, and $A_{\rm G}=0.223$~mag, we derived $M_{\rm G}=1.66$~mag, and a system luminosity of 17.09~L$_{\sun}$, assuming zero bolometric correction. However, the system's luminosity and distance value obtained from the $V$-band LC+RV solution (28.17~L$_{\sun}$ and $537\pm42$~pc) do not match the values given in the Gaia~DR3 catalogue (37.32 or 17.09~L$_{\sun}$ and $409\pm3$~pc). It should be noted that the interstellar absorption values from the FLAME module and the ESP-HS module in the Gaia~DR3 catalogue are quite different (0.789 and 0.223~mag, respectively), and similarly, the RR~Lep system's luminosities also show a large difference. Furthermore, there is still a difference of approximately 100~pc between the distance calculated from our $V$-band LC+RV solution and that from the Gaia~DR3 parallax. If we substitute the interstellar absorption used in our calculations with the value of 0.8~mag, as calculated from the FLAME module, this difference is eliminated. We conclude that the aforementioned discrepancies arise due to the strength of the tertiary component’s spectral lines, which were blended with those of the primary component at most phases (Sect.~\ref{Sec:Spectroscopy}). Around the first quadrature, at phases 0.17 and 0.25, spectra of the tertiary component were clearly evident as its sodium~D lines were considerably stronger than those of the primary star.

\begin{figure}[t]
\centering
\includegraphics[width=\columnwidth]{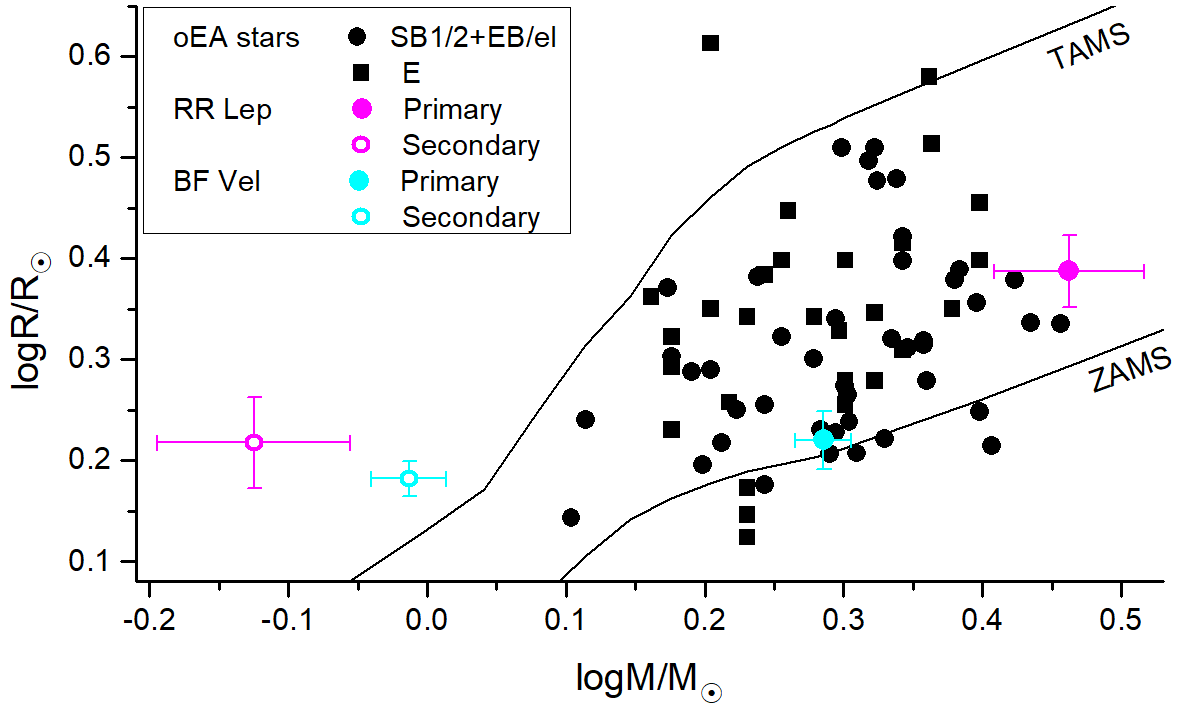}  \\
\includegraphics[width=\columnwidth]{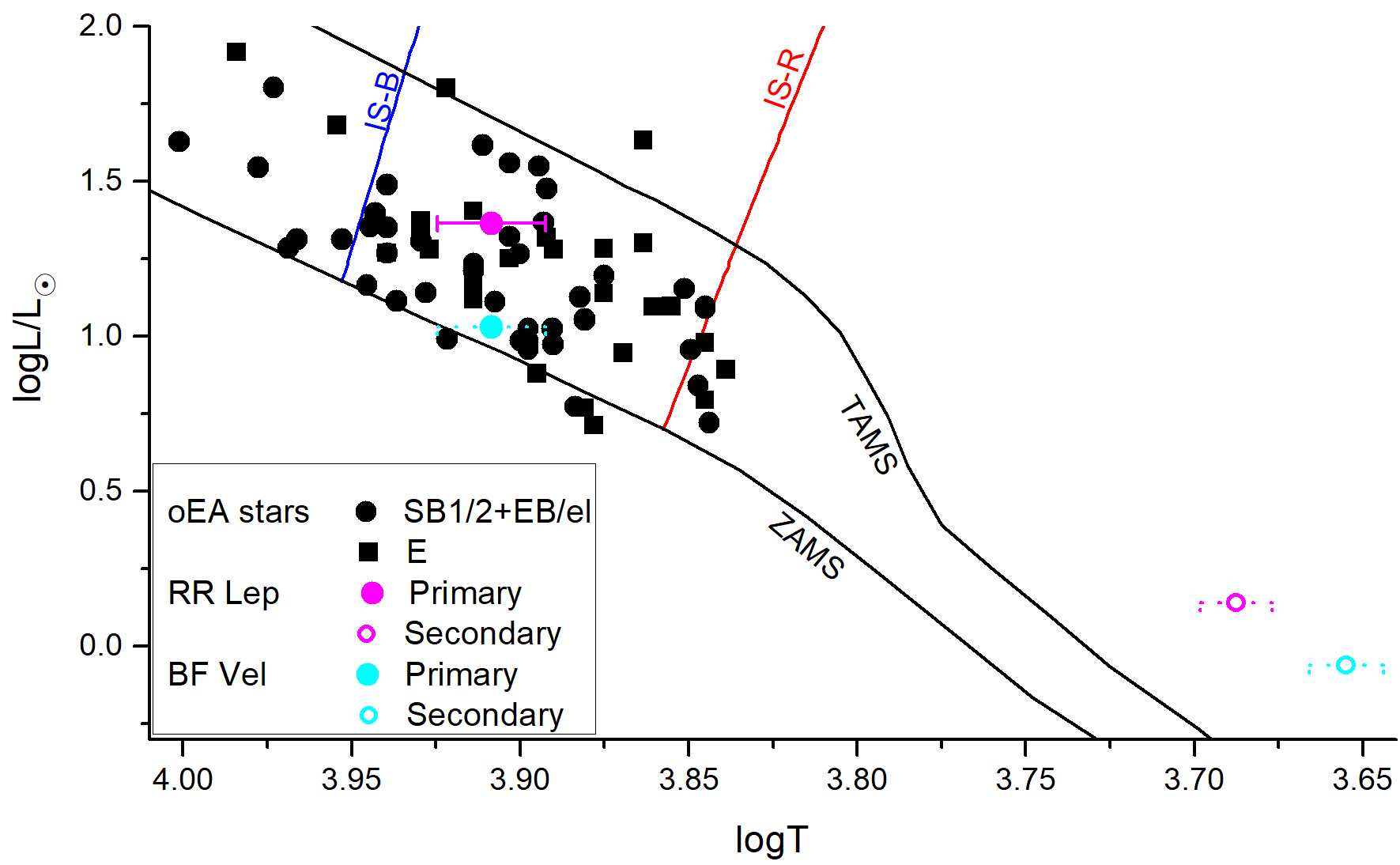}
\caption{Mass--Radius (top panel) and Hertzsprung--Russell (bottom panel) evolutionary diagrams. In both diagrams the black symbols denote the $\delta$~Scuti members of oEA stars that belong either to spectroscopic single- or double-lined (SB1 or SB2) and eclipsing (EB) or ellipsoidal (el) systems (dots) or only to eclipsing systems (squares). The sample was mainly gathered from \citet{LIAN17}. Magenta and cyan symbols refer to the primary (filled) and secondary (open) components of RR~Lep and BF~Vel, respectively. Solid black lines indicate the ZAMS and TAMS limits and the coloured solid lines (B = blue, R = red) the boundaries of the instability strip \citep[IS;][]{SOY06b}.}
\label{fig:EVOL}
\end{figure}

The physical properties of the primary components of these systems and their pulsational behaviour (Sect.~\ref{Sec:Puls}) indicate that they are $\delta$~Sct stars. In order to compare their properties with similar cases, we used mostly the sample of \citet{LIAN17} for the oEA stars and various works for individual cases from the literature. The gathered sample contains 96 oEA systems (including BF~Vel and RR~Lep), however there is information about the absolute parameters of only 72 of these. In particular, 33 of them are double-lined spectroscopic and eclipsing systems (SB2+EB), which are the most important tools for calculating absolute parameters of stars. Nine of them are SB1+EB systems and one is SB1+ellipsoidal system (SB1+el); the latter systems can be considered as the second most informative cases. The masses of the pulsating stars in the other 29 oEA systems were assumed, based on their spectral type. The locations of the components of RR~Lep and BF~Vel in the Mass-Radius (M--R) and Hertzsprung-Russell (H--R) evolutionary diagrams are given in Fig.~\ref{fig:EVOL}.
The purpose of these diagrams is to a) check the evolutionary stages of both components of each studied system and b) compare the evolutionary stages of their primaries with other $\delta$~Sct stars in oEA systems. For the latter reason, we included only the pulsating members of the oEA stars and not their non-pulsating companions.

The position of the primary component of RR~Lep in the M--R diagram indicates that it is well inside the main-sequence boundaries and is identified as the most massive $\delta$~Sct star of the current sample. The primary of BF~Vel is located on the ZAMS boundary in the same diagram. The positions of the secondary components in both diagrams show that they have evolved out of the main-sequence. The primary components lie well inside the classical instability strip as shown in the H--R diagram. As both BF~Vel and RR~Lep are semi-detached systems, their primary stars are accreting mass from their companions, which have evolved to the sub-giant stage of evolution.

\section{Eclipse Timing Variation analysis}
\label{Sec:ETV}	

For the analysis of orbital period variations we employed the `LITE' code of \citet{ZAS09}. The code uses statistical weights ($w$) on the times of minima (ToM) based on the method followed for the observations of the eclipses. Older ToM derived from visual or photographic data were assigned $w=1$, the photoelectric $w=7$, and the CCD $w=10$. Past ToM of each studied system were obtained from online databases\footnote{https://www.variablestarssouth.org/}$^,$\footnote{http://var2.astro.cz/ocgate/}$^,$\footnote{http://www.oa.uj.edu.pl/ktt/krttk\_dn.html}$^,$\footnote{https://www.bav-astro.eu/index.php/}$^,$\footnote{https://www.as.up.krakow.pl/ephem/} and were enriched with those derived from our ground-based data, TESS data and data from other ground-based surveys. The ToM used in the following analyses are given in Table~\ref{Tab:ToM}.

The code is able to fit the data points with Light-Time Effect \citep[LITE;][]{IRW59} and parabolic curves. The LITE curve has seven free parameters, namely the ephemeris of the EB ($T_0$ and $P$), the amplitude of the variation $A$, and the orbital period $P_3$, the time of the periastron passage $JD_0$, the argument of periastron $\Omega$, and the orbital eccentricity $e_3$ of the tertiary component. The parabolic fitting has three free parameters; the ephemeris of the binary and the parabolic coefficient $C_2$. We used ephemerides from \citet{KRE04} for both systems; they were left free to adjust during the iterations. The selection of the final model is based also on the Bayesian information criterion (BIC). The results of the Eclipse-Timing Variation (ETV) analyses for both systems were applied as inputs in the `InPeVeb' software \citep{LIA15} to calculate the parameters of the most likely orbital modulating mechanisms.

\subsection{RR~Lep}
\label{Sec:ETV_RRLep}

The ETV analysis of RR~Lep is based on 140 ToM spanning from the mid 1930's up to 2023. A periodic variation due to a tertiary companion is not particularly obvious in the ETV diagram. However, as a noticeable third light was detected in both the spectroscopic and photometric observations, a LITE curve was selected for fitting the ToM. Given that the system has a semi-detached configuration, a parabola was also tested as a possible solution.

\begin{figure}
\centering
\includegraphics[width=8.5cm]{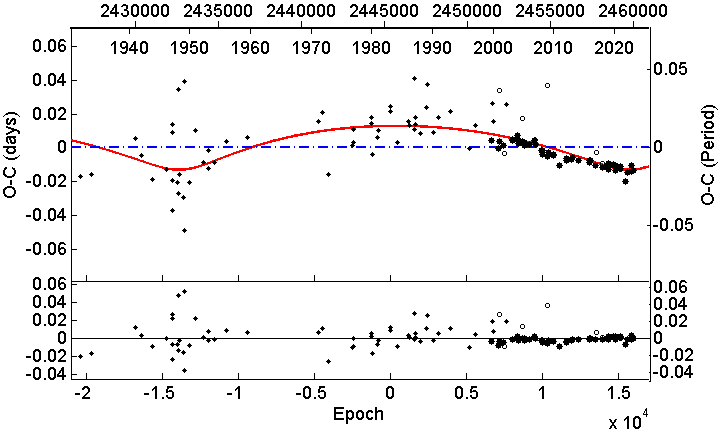}  \\
\includegraphics[width=8.5cm]{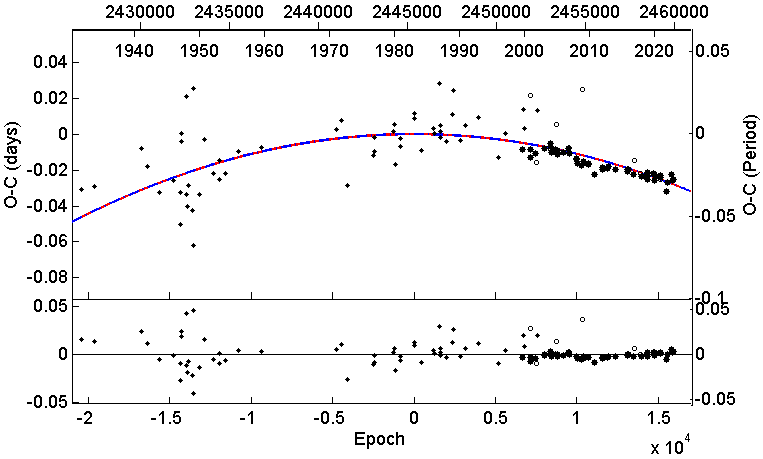}  \\
\caption{Fitting of a LITE curve (upper plot) and of a downward parabola (lower plot) on the ToM of RR~Lep. The bigger the symbol the bigger the statistical weight. Open and filled symbols denote the primary and secondary ToM, respectively. Lower panels show the residuals of each solution.}
\label{fig:LEPRRETV}
\vspace{5mm}
\centering
\includegraphics[width=8.5cm]{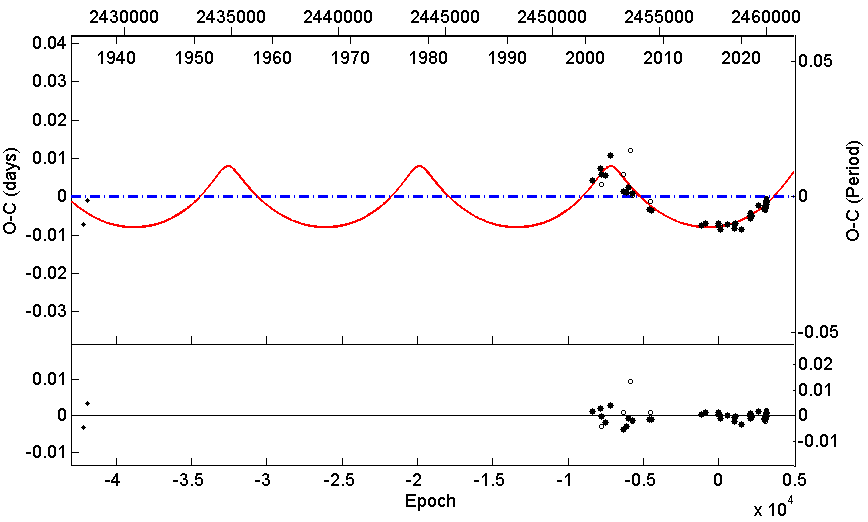} \\
\includegraphics[width=8.5cm]{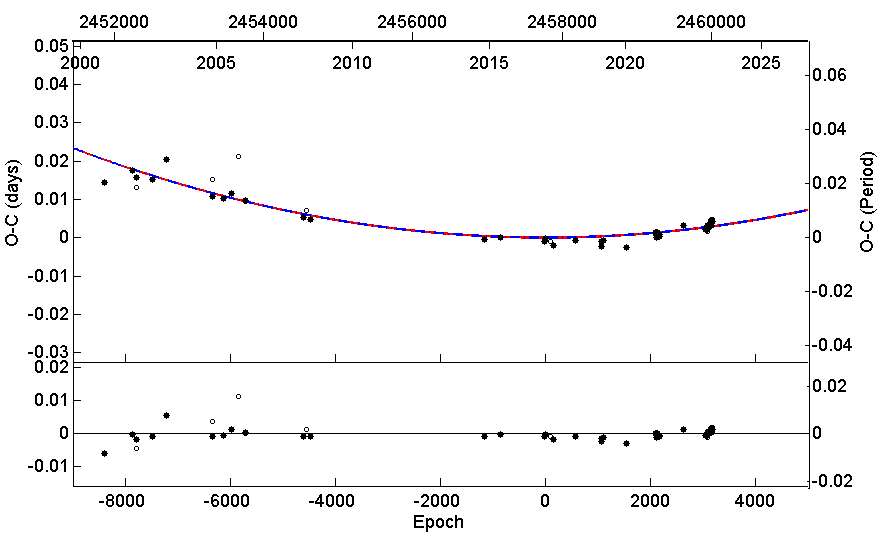}
\caption{Fitting of a LITE curve (upper plot) and of an upward parabola (lower plot) on the ToM of BF~Vel. Symbols have the same meaning as in Fig.~\ref{fig:LEPRRETV}.}
\label{fig:VELBFETV}
\end{figure}

The LITE solution is plotted in the upper panel of Fig.~\ref{fig:LEPRRETV} and its parameters are listed in Table~\ref{Tab:ETVLITE}. This solution suggests that a tertiary component in eccentric orbit with a period of $\sim 74$~yr orbits the EB. The mass function of the third body suggests a minimum mass of $\sim 0.33$~M$_{\sun}$ (i.e. in coplanar orbit with the EB). However, following the formalism of \citet{LIAN12}, we found that if the tertiary member is a main-sequence star, then, in order to contribute $\sim 6\%$ of the total luminosity (see Table~\ref{Tab:Models_RRLep}), its mass should be $\sim 1.14$~M$_{\sun}$. The latter is possible if its inclination is $\sim 19.5\degr$ with respect to the EB's orbital plane. Alternatively, a higher inclination angle of the third body's orbit, which would yield a lower mass value, is also possible under the assumption that the tertiary component is an evolved star (i.e. subgiant or giant).

The parabolic fitting resulted in a downward parabola (lower panel of Fig.~\ref{fig:LEPRRETV}) that suggests an orbital period decrease. This contradicts the conventional semi-detached configuration (i.e. the less massive star transfers mass to the more massive component). However, the derived curvature of the parabola may indicate non-conservative mass transfer ($\dot{M}_{\rm tr}$) and mass loss ($\dot{M}_{\rm loss}$) due to magnetic braking \citep{HIL01, ERD05}. Therefore using the method of \citet{ERD05}, we assumed that the secondary component is the mass donor star of the system and transfers mass with a typical rate of $10^{-8}$~M$_{\sun}$~yr$^{-1}$ to the primary, while simultaneously losing mass due to magnetic braking. The gyration constant of this component was assumed as $k=0.32$, while its Alfv\'{e}n radius as 10$R_2$ \citep[cf.][]{SOY11, LIA13}. Results regarding this solution are given in Table~\ref{Tab:ETVPar}.

The BIC values for the proposed solutions resulted as $-1075$ for the LITE and $-1073$ for the parabola, indicating that LITE is more appropriate solution. Moreover, as our photometric and spectroscopic data demonstrate the presence of a tertiary component, and a parabolic fitting in the ToM of RR~Lep needs many assumptions to explain the ETV, we also conclude that the correct solution for the observed orbital period changes of this system is the LITE.

\begin{table}					
\begin{center}					
\caption{LITE parameters of the ETV analysis of RR~Lep and BF~Vel.}					
\label{Tab:ETVLITE}
\scalebox{0.95}{
\begin{tabular}{lcc}					
\hline					
System:	&	RR~Lep	&	BF~Vel	\\
\hline					
Parameter	&		&		\\
\hline					
	&	\multicolumn{2}{c}{Eclipsing binary}			\\
\hline					
$T_0$ (HJD)	&	2445352.632$\pm$0.002	&	2457777.167$\pm$0.001	\\
$P$ (d)	&	0.9154267$\pm$0.0000002	&	0.7040275$\pm$0.0000001	\\
\hline					
	&	\multicolumn{2}{c}{Tertiary component}			\\
\hline					
$P_3$ (d)	&	74.2$\pm$0.6	&	24.5$\pm$0.3	\\
$JD_0$ (HJD)	&	 2459956$\pm$1235	&	2434825$\pm$332	\\
$A$ (d)	&	0.013$\pm$0.002	&	0.008$\pm$0.001	\\
$\Omega$ ($\degr$)	&	277$\pm$39	&	87$\pm$11	\\
e	&	0.53$\pm$0.25	&	0.63$\pm$0.06	\\
$f(m_3)$ (M$_{\sun}$)	&	0.0024$\pm$0.0002	&	0.0044$\pm$0.0001	\\
\hline					
$\sum res^2$	&	0.051	&	0.0019	\\
\hline					
\end{tabular}}	
\end{center}					
\begin{center}					
\caption{Mass transfer/loss parameters for both systems as alternative explanations for their ETVs.}					
\label{Tab:ETVPar}
\scalebox{0.9}{					
\begin{tabular}{lcc}					
\hline					
System:	&	RR~Lep	&	BF~Vel	\\
\hline					
Parameter	&		&		\\
\hline					
$T_0$ (HJD)	&	2445352.644$\pm$0.001	&	2457777.161$\pm$0.001	\\
$P$ (d)	&	0.9154267$\pm$0.0000001	&	0.7040279$\pm$0.0000002	\\
$c_2$ (d cycle$^{-1}$)	&	-1.1$\pm0.1~\times 10^{-10}$	&	2.9$\pm0.3~\times 10^{-10}$	\\
$\dot{P}$ (d~yr$^{-1}$)	&	-8.8$\pm0.8~\times 10^{-8}$	&	3.0$\pm0.3~\times 10^{-7}$	\\
$\dot{M}_{\rm tr}$ (M$_{\sun}$~yr$^{-1}$)	&	$10^{-8,~~\rm a}$	&	2.8$\pm0.4~\times 10^{-7}$	\\
$\dot{M}_{\rm loss}$ (M$_{\sun}$~yr$^{-1}$)	&	-3.4$\pm0.5~\times 10^{-8}$	&	--	\\
\hline					
$\sum res^2$	&	0.059	&	0.0029	\\
\hline	
$^{\rm a}$assumed
\end{tabular}}							
\end{center}					
\end{table}

\subsection{BF~Vel}
\label{Sec:ETV_BFVel}

There are 61 ToM currently available for BF~Vel. All of them, except for two, cover the time period after the year 2000; the other two are from the late 1930s and were based on photographic plates. Taking these old ToM into account, we fitted a LITE curve to the data points. The selection of this curve is based on the detection of third light in both spectroscopic and photometric observations. The fitted curve is illustrated in the upper plot of Fig.~\ref{fig:VELBFETV} and its parameters are listed in Table~\ref{Tab:ETVLITE}. According to this solution, a third body orbits the EB with a period of $\sim 25$~yr. The $f(m_3)$ of this solution derives a minimum mass of $\sim 0.36$~M$_{\sun}$. Therefore, similar to that discussed above for RR~Lep, its inclination with respect to the EB's orbit should be $\sim24\degr$ in order for its mass as a dwarf to be $\sim1$~M$_{\sun}$, which is sufficient to explain the observed $\sim 8\%$ contribution in the system's total luminosity (see Table~\ref{Tab:Models_BFVel}). Alternatively, a less massive, but more evolved star (i.e. more luminous) at a higher inclination angle, is also possible. It should be mentioned that a $P_3$ of $\sim 36.75$~yr is also possible. The fit on the data points using this value yielded slightly larger $\sum res^2$ (i.e.~0.0022) but also a much smaller $f(m_3)$ value (i.e.~0.0021~M$_{\sun}$) that cannot easily explain the observed light contribution from the tertiary component. Therefore we neglected this solution.

As a second solution, the data points could be fitted by an upward parabola, which agrees with the semi-detached status of the system. For this case, the fitting is given in the lower plot of Fig.~\ref{fig:VELBFETV} and the corresponding parameters in Table~\ref{Tab:ETVPar}. However, that does not accommodate the presence of the tertiary component. The BIC value of LITE ($-604$) is lower than that of the parabolic fitting ($-573$). Therefore, the LITE curve is the better solution, not only statistically, but also because it explains the presence of the tertiary component.

\section{Pulsation Analyses}
\label{Sec:Puls}

The frequency search for both systems relied only on the TESS LC residuals because of their photometric accuracy, time resolution and continuity in time (i.e. no time gaps). Based on the spectral types and the physical properties of the components of both systems, as described in the previous sections, only the primary stars have properties of $\delta$~Scuti stars. Therefore, the data points between orbital phases 0.9--0.1, when the pulsating components are eclipsed, were excluded from the frequency analyses. The software PERIOD04 \citep[classical Fourier analysis;][]{LEN05} was used to extract the pulsation frequencies. Although the typical oscillation range for $\delta$~Scuti stars is 4-80~d$^{-1}$ \citep{BRE00, BOW18}, we extended the search from 0 to 80~d$^{-1}$ in order to detect possible $\delta$~Scuti--$\gamma$~Doradus hybrid behaviour, or low-frequency modes that are connected to $g$-mode pulsations or the orbital frequency ($f_{\rm orb}$) or other frequency combinations. The noise level of each detected frequency was computed around it with a spacing of 2~d$^{-1}$ and a box size of 2 \citep{BAL14}. After the computation of a frequency, the residuals were pre-whitened for the next one. The search was stopped when the signal-to-noise ratio of the detected frequency was approximately 5 \citep[S/N$\sim5$;][]{BAR15, BOW21}. The TESS data sets used for the analyses were observed in short-cadence mode (i.e. 2-min time resolution) and span over approximately 25 continuous days. The Nyquist frequency for these data sets is $\sim 360$~d$^{-1}$ and the frequency resolution \citep[$1.5/T$, where $T$ is the observations time range in days; cf.][]{LOU78} is $\sim 0.06$~d$^{-1}$.

The results of the analyses are divided into two main parts; the independent frequencies and the combination frequencies. In addition to the Fourier modelling of the pulsations, we also estimated the most likely $l$-degrees of the independent frequencies. For this, we used the method of \citet{BRE00} for calculating the pulsation constant $Q$ of each frequency \citep[see also][]{LIA22} by using the advantage of the known absolute properties of the pulsating star (Sect.~\ref{Sec:AbsPar}). Then, we compared the $Q$ values with the models of \citet{FIT81} in order to derive the most likely oscillation modes. Table~\ref{Tab:FreqsInd} includes the results only for the independent frequencies detected in each system. In particular, we list: the increasing number of the frequency ($n$), its value ($f_{\rm n}$), amplitude ($A$), phase ($\Phi$), S/N, $Q$, and $l$~degrees. Table~\ref{Tab:FreqsDep} contains the same information as Table~\ref{Tab:FreqsInd}, except the $Q$ and the $l$-degrees, but for the combination frequencies.

\begin{table*}
\begin{center}															
\caption{Independent oscillation frequencies of the primary components of RR~Lep and BF~Vel based on the TESS data sets.}									
\label{Tab:FreqsInd}	
\begin{tabular}{cccc cccc}															
\hline															
$n$	&	  $f_{\rm n}$	&	$A$	&	  $\Phi$	&	S/N	&	$Q$	&	$l$-degree	&	mode	\\
	&	     (d$^{-1}$)	&	(mmag)	&	($2\pi$~rad)	&		&	(d)	&		&		\\
\hline															
\multicolumn{8}{c}{RR~Lep}															\\
\hline															
1	&	32.2753$\pm$0.0001	&	3.01$\pm$0.01	&	0.544$\pm$0.001	&	34.7	&	0.014$\pm$0.002	&	3	&	p4	\\
2	&	31.8665$\pm$0.0001	&	2.66$\pm$0.01	&	0.144$\pm$0.001	&	32.9	&	0.014$\pm$0.002	&	3	&	p4	\\
5	&	23.6632$\pm$0.0002	&	1.29$\pm$0.01	&	0.220$\pm$0.002	&	24.2	&	0.019$\pm$0.002	&	1 or 0	&	p3 or R	\\
6	&	30.1510$\pm$0.0002	&	1.24$\pm$0.01	&	0.490$\pm$0.002	&	13.0	&	0.015$\pm$0.002	&	2	&	p4	\\
7	&	32.4916$\pm$0.0003	&	1.09$\pm$0.01	&	0.658$\pm$0.002	&	12.0	&	0.014$\pm$0.002	&	3	&	p4	\\
\hline															
\multicolumn{8}{c}{BF~Vel}															\\
\hline															
3	&	46.7311$\pm$0.0007	&	0.79$\pm$0.02	&	0.812$\pm$0.005	&	16.4	&	0.014$\pm$0.002	&	3 or 0	&	p4 or R4H	\\
\hline															
\end{tabular}															
\end{center}															
\end{table*}

\subsection{RR~Lep}
\label{Sec:Puls_RRLep}

Two TESS time series are currently available for RR~Lep, namely Sector~5 and 32, and they span over 25.6~d. The data set of Sector~5 has a time resolution of 2~minutes and that of Sector~32, 10~minutes. Furthermore, these data sets were acquired with a separation of 1.94~years. Therefore only the data of Sector~5 were used (app.~13800 data points) in the analysis to minimise introduction of alias frequencies \citep{BRE00}.

\begin{figure}
\centering
\includegraphics[width=8.9cm]{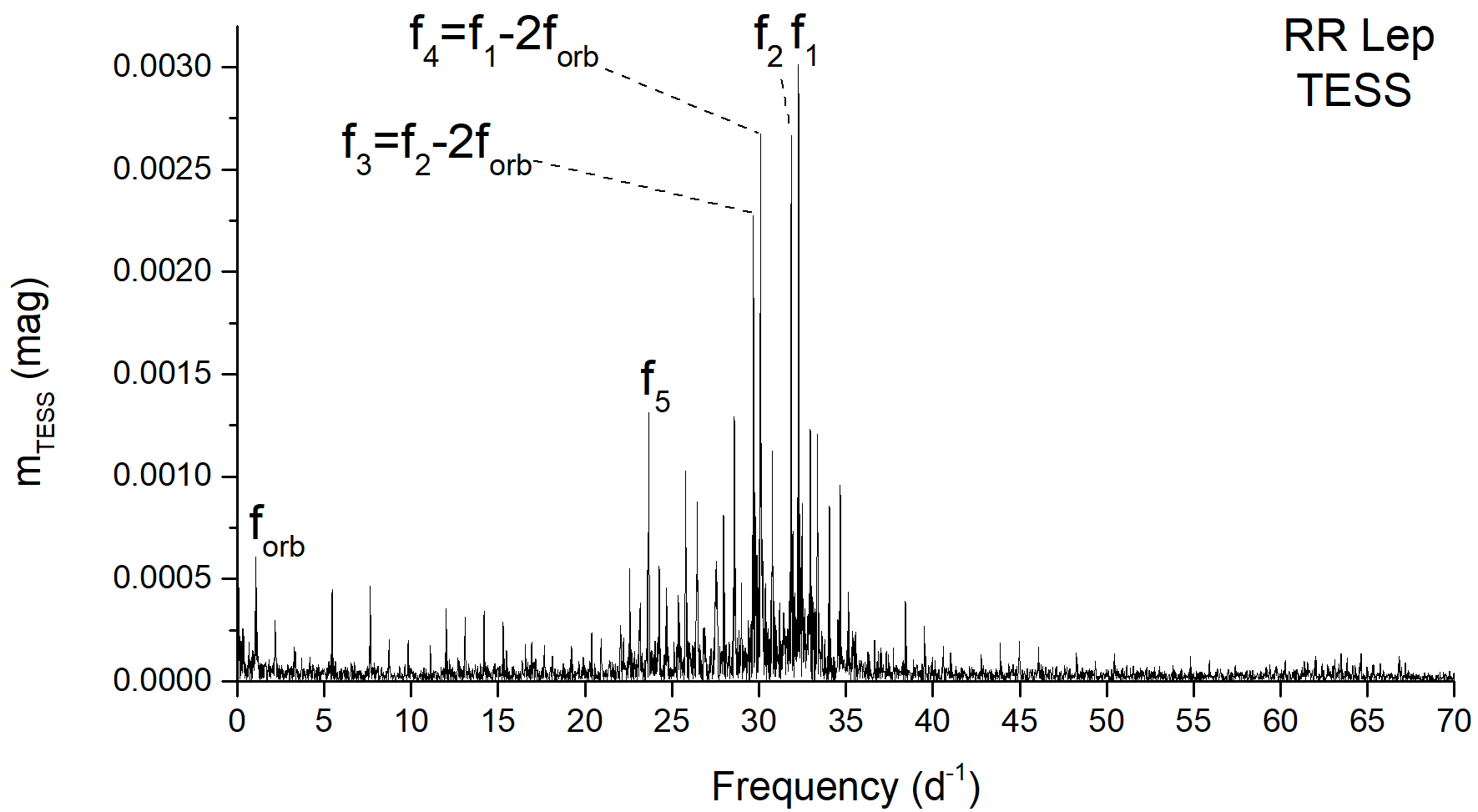}  \\
\includegraphics[width=8.9cm]{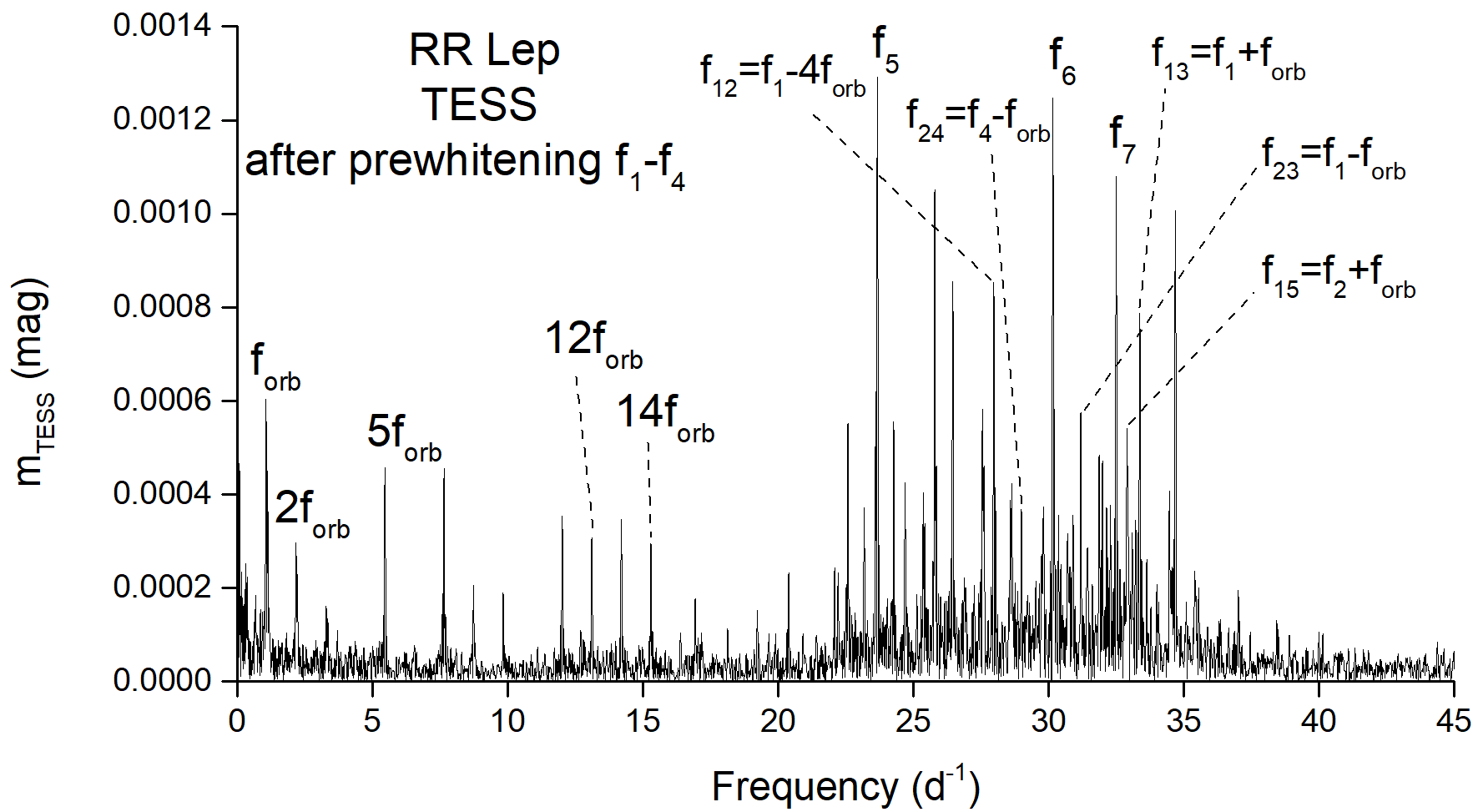}
\caption{Periodogram of the pulsating component of RR~Lep. The strongest frequencies are indicated.}
\label{fig:RRLFS}
\end{figure}

The pulsating member of RR~Lep was found to oscillate in more than 30 frequencies. However, only five, namely $f_1$, $f_2$, $f_5$, $f_6$, and $f_7$, are identified as independent. We detected many harmonics of the $f_{\rm orb}$ (=1.0924~d$^{-1}$). The other frequencies are combinations of others and many of them result from the rotational splitting of the independent frequencies by the $f_{\rm orb}$ and its harmonics due to the orbital motion of the pulsator around the common centre of mass. The rotationally split modes would not be spaced exactly equally by the orbital frequency due to the influence of the Coriolis force on the pulsations \citep{LED51}. According to the $Q$ values of the independent frequencies, all of them are identified as non-radial pressure modes with their $l$-degrees to range between 1 and 3. The ratio $P_{\rm pul}/ P_{\rm orb}$ for these frequencies is lower than 0.046, which agrees with the conclusion of \citet[][i.e. $P_{\rm pul}/P_{\rm orb}<0.07$]{ZHA13} regarding the non-radial nature of these frequencies. On the other hand, we note that $f_5/f_1 \sim f_5/f_7 \sim 0.73$, $f_5/f_2 \sim 0.74$, and $f_5/f_6 \sim 0.78$. These frequency ratios are typical for the radial fundamental and the first overtone modes \citep{STE79, OAS06, AER10}. Therefore, $f_5$ could be a radial fundamental mode. However due to the very close values of the other independent frequencies, we cannot be certain which of them could be the first overtone. The periodogram of RR~Lep, including some of the strongest detected frequencies, is shown in two plots with different scales in Fig.~\ref{fig:RRLFS}, and a representative Fourier fit on the data points is illustrated in Fig.~\ref{fig:FF}. Detailed results for the frequency search are given in Tables~\ref{Tab:FreqsInd} and \ref{Tab:FreqsDep}.

The comparison of the results of \citet{LIA13} with the present ones shows that the authors of that work had found as the most dominant frequency the $+1$ alias of the present $f_1$. Hence, for reasons of completeness, we also re-analysed the ground-based $B$ and $V$ residual data of \citet{LIA13} based on the LC+RV model presented herein (Sect.~\ref{Sec:Models}). The scope of this re-analysis was the determination of the oscillation modes using the amplitude ratio and the phase difference of a given frequency value from two different photometric filters. The new analysis of these data yielded the same alias frequency. Therefore, we fixed it at the value found from the analysis of TESS data and we tried to detect other frequencies as well. Unfortunately, given that the sample of data points and the photometric accuracy of these data sets are rather inferior in comparison to that of TESS, in addition to the dominant frequency only a second one, which differs between the filters, with S/N>4 (Table~\ref{table:RRLepFreqsInd}) was detected. The mode identification for the dominant frequency was made using the FAMIAS software \citep{ZIM08} and the MAD-$\delta$~Sct models \citep{MON07}. Particularly, the closest model to the stellar and pulsational parameters of the primary of RR~Lep was found for a star with a mass of 2.2~M$_{\sun}$ (i.e. the maximum value in the MAD models), a temperature of 8470~K, and $\log g=3.94$~cm~s$^{-2}$. This model suggests an $l$-degree equal to 3, similarly to that found for $f_1$ of the TESS data (Table~\ref{Tab:FreqsInd}).

\subsection{BF~Vel}
\label{Sec:Puls_BFVel}

The frequency search of BF~Vel was based only on the short-cadence (2-min time resolution) TESS data set of Sector~35. This data set spans over approximately 24~days, but there is a 6-day time gap. The total number of points used for the analysis, after the exclusion of the primary eclipses, is 10589. Unfortunately, our ground-based photometric data cover much shorter time spans (i.e.~$\sim$2.5~d in $B$ and $\sim$4~d in $V$ and $I$ filters) and have much lower photometric accuracy, therefore, they were excluded from the analysis. The TESS data set of Sector~62, although it has the same time resolution and covers about the same time span as that of Sector~35, it includes $\sim30\%$ fewer data points and has a time difference of approximately two years with that of Sector~35. Therefore, as the combination of these data sets into one common frequency analysis is subject to alias, we excluded this data set from our analyses.

\begin{figure}
\centering
\includegraphics[width=8.9cm]{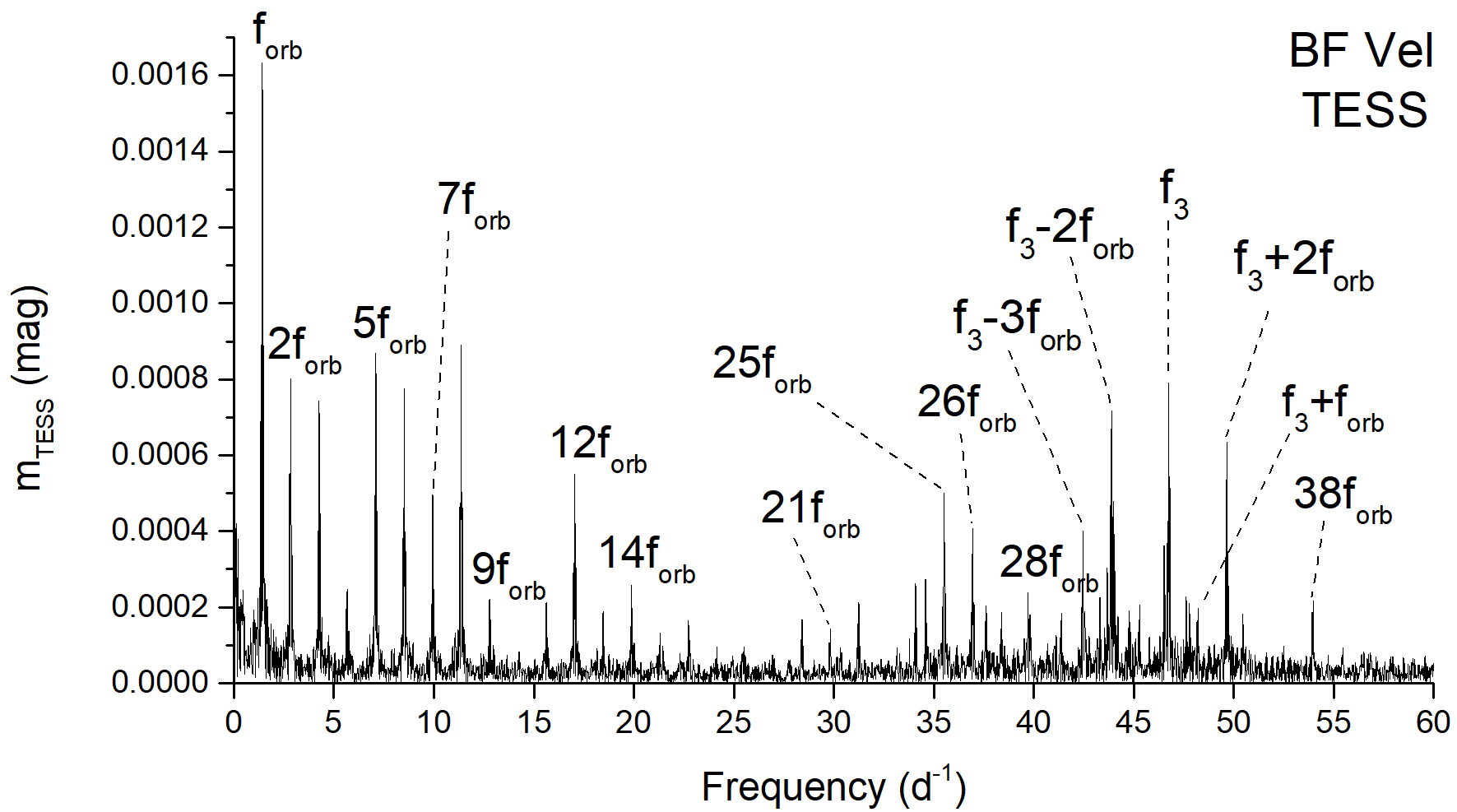}
\caption{The same as Fig.~\ref{fig:RRLFS} but for BF~Vel.}
\label{fig:BFVFS}
\end{figure}

\begin{figure*}
\centering
\includegraphics[width=18cm]{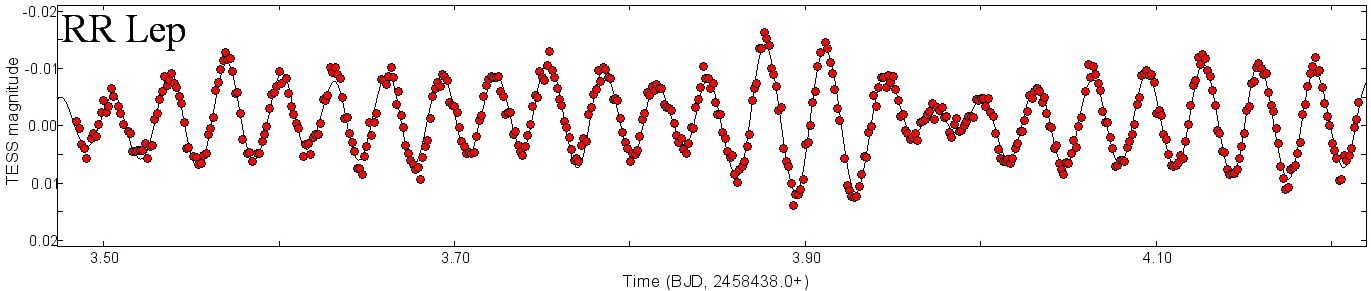}  \\
\includegraphics[width=18cm]{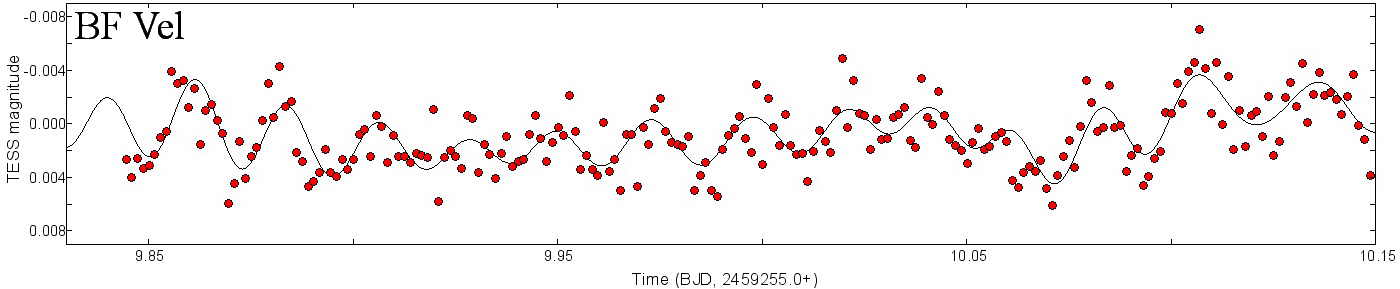}
\caption{Fourier fittings on a small fraction of the available TESS data for each system studied.}
\label{fig:FF}
\end{figure*}

The periodogram of the oscillating member of BF~Vel (Fig.~\ref{fig:BFVFS}) is dominated by the orbital frequency and its harmonics. We detected most of the harmonics of $f_{\rm orb}$ up to 38$f_{\rm orb}$. Only one independent frequency was detected, that is $f_3=46.73$~d$^{-1}$, and also a few frequencies that indicate rotational splitting by the $f_{\rm orb}$ and its harmonics. In total, 37 oscillation frequencies were detected, but we cannot be certain of the intrinsic origin of all of them, since many of them were found to be connected to an extremely low-frequency value (i.e.~$f_6$) or to high-order multiples of the $f_{\rm orb}$. The $Q$ value of $f_3$, based on the absolute parameters of the primary component and the same method as in the case of RR~Lep, was found to be $0.014\pm0.002$~d. According to models of \citet{FIT81}, $f_3$ is either a $p$-mode or a radial mode and particularly the 4$^{\rm th}$ overtone of the radial fundamental mode. The pulsations model is given in Tables~\ref{Tab:FreqsInd} and \ref{Tab:FreqsDep}, while a sample of the Fourier fitting is presented in Fig.~\ref{fig:FF}.
		
\begin{table}									
\begin{center}
\caption{Oscillation frequencies of the primary component of RR~Lep using the $B$ and $V$ ground-based data of \citet{LIA13}.}									
\label{table:RRLepFreqsInd}									
\begin{tabular}{ccccc}									
\hline									
$n$	&	  $f_{\rm n}$	&	$A$	&	  $\Phi$	&	S/N	\\
	&	     (d$^{-1}$)	&	(mmag)	&	($2\pi$~rad)	&		\\
\hline									
\multicolumn{5}{c}{$B$~filter}									\\
\hline									
1	&	32.2787$\pm$0.0004	&	9.8$\pm$0.4	&	0.567$\pm$0.006	&	7.8	\\
2	&	25.5633$\pm$0.0009	&	3.89$\pm$0.4	&	0.831$\pm$0.015	&	4.0	\\
\hline									
\multicolumn{5}{c}{$V$~filter}									\\
\hline									
1	&	32.2768$\pm$0.0005	&	7.76$\pm$0.4	&	0.583$\pm$0.008	&	6.3	\\
2	&	27.0255$\pm$0.0010	&	3.80$\pm$0.4	&	0.027$\pm$0.017	&	4.0	\\
\hline									
\end{tabular}									
\end{center}									
\end{table}

\section{Summary, discussion, and conclusions}
\label{Sec:Disc}

Multi-epoch, multi-filtered ground-based and high-accuracy space-borne photometry of the systems RR~Lep and BF~Vel have been combined with medium resolution spectroscopy to derive the physical properties of their components and, furthermore, to study their pulsational behaviour. Both systems are identified as oEA stars with their primary components displaying $\delta$~Sct type pulsations. Radial velocities of both components of the binary pairs were determined with broadening function analyses of the blue spectra. These analyses also indicated the presence of tertiary companions, which were confirmed with spectra of H$_{\alpha}$ and Na~I~D lines. Orbital period modulations were analysed to determine probable orbital period modulating mechanisms. The secondary stars of both systems have evolved to the subgiant stage, and are chromospherically active.

Although for both systems a significant contribution of third light was detected in the photometric and spectroscopic observations and analyses, they lack clear periodic variations in their orbital periods that could be attributed to the existence of their tertiary companions. However, we conclude that the LITE is the most appropriate orbital period modulating mechanism.

The primary star of RR~Lep pulsates in five independent frequencies, with the dominant being 32.275~d$^{-1}$. All these frequencies are identified as $p$-modes, but $f_5$ could also be a radial fundamental mode. In addition, 31 combination frequencies were detected, many of which are connected to the orbital frequency of the system. This star is the most massive (2.9~M$_{\sun}$) among the $\delta$~Sct stars in the sample of the oEA systems. Our results regarding the pulsation modelling partially agree with those of \citet{KAH24}. They found seven independent frequencies with the most dominant one to be 31.866~d$^{-1}$ followed by the 32.276~d$^{-1}$ as the second strongest one. We also detected these two frequencies ($f_1$ and $f_2$) but in reverse order (i.e. the amplitude of $f_1$ was found larger than that of $f_2$). The frequencies 23.665~d$^{-1}$ and 32.492~d$^{-1}$ were detected in both studies ($f_5$ and $f_7$ in our work). The main discrepancies between the two works concern the current $f_6$, which was not detected by \citet{KAH24}, and their classification for the 25.782~d$^{-1}$, 26.449~d$^{-1}$, and 31.986~d$^{-1}$. The latter frequencies were also detected in our work (i.e. $f_8$, $f_9$, and $f_{21}$), but we found that they are just combinations of other frequencies. Larger discrepancies occur between our work and that of \citet{LIA13}, who detected only two frequencies, namely 33.280~d$^{-1}$ and 24.318~d$^{-1}$. The first one in their study is probably the alias of the present $f_1$ (i.e. $f_1=33.280-1$~d$^{-1}$), which could be caused by the time gaps in the ground-based LCs that they used \citep[alias frequencies, cf.][]{BRE00}. Although we re-analysed the data of \citet{LIA13} by setting the dominant frequency $f_1$ to that found from the TESS data, we did not detect any other frequencies in the $B$ and $V$ ground-based data sets similar to those detected in the TESS data. However, for $f_1$, both methods of mode identification (i.e. models of \cite{FIT81} and the MAD models) yielded the same $l$-degree value ($l$=3). The TESS data that we used for modelling is far more precise than that used in previous studies with ground-based data; however, the latter are useful for checking the main results, not for pulsation modelling.


One independent frequency mode (=46.731~d$^{-1}$) was found for the $\delta$~Sct star of BF~Vel, and more than 35 other combination frequencies were detected. The majority of these frequencies are splittings of the orbital frequency due to rotation of the star around the systemic centre of mass, but many of them are probably leftovers from other stronger frequencies. Our results have large discrepancies compared with those of \citet{MAN09}. In particular, they detected seven frequencies with the dominant one being 44.94~d$^{-1}$, which is close to our $f_3$. The other frequencies reported in that work are either harmonics of others or they were not characterized at all. With the exception of their 7.48~d$^{-1}$, they did not detect any other slow frequencies, in contrast to our results. However, it should be noted that the results of \citet{MAN09} were based on a much smaller data sample (i.e.~350 points in $B$~filter spanning over three consecutive nights) in comparison with the TESS data that we used. The short time span of their data series could also be the reason they did not detect slow frequencies connected to the $f_{\rm orb}$.

\begin{figure}
\centering
\includegraphics[width=\columnwidth]{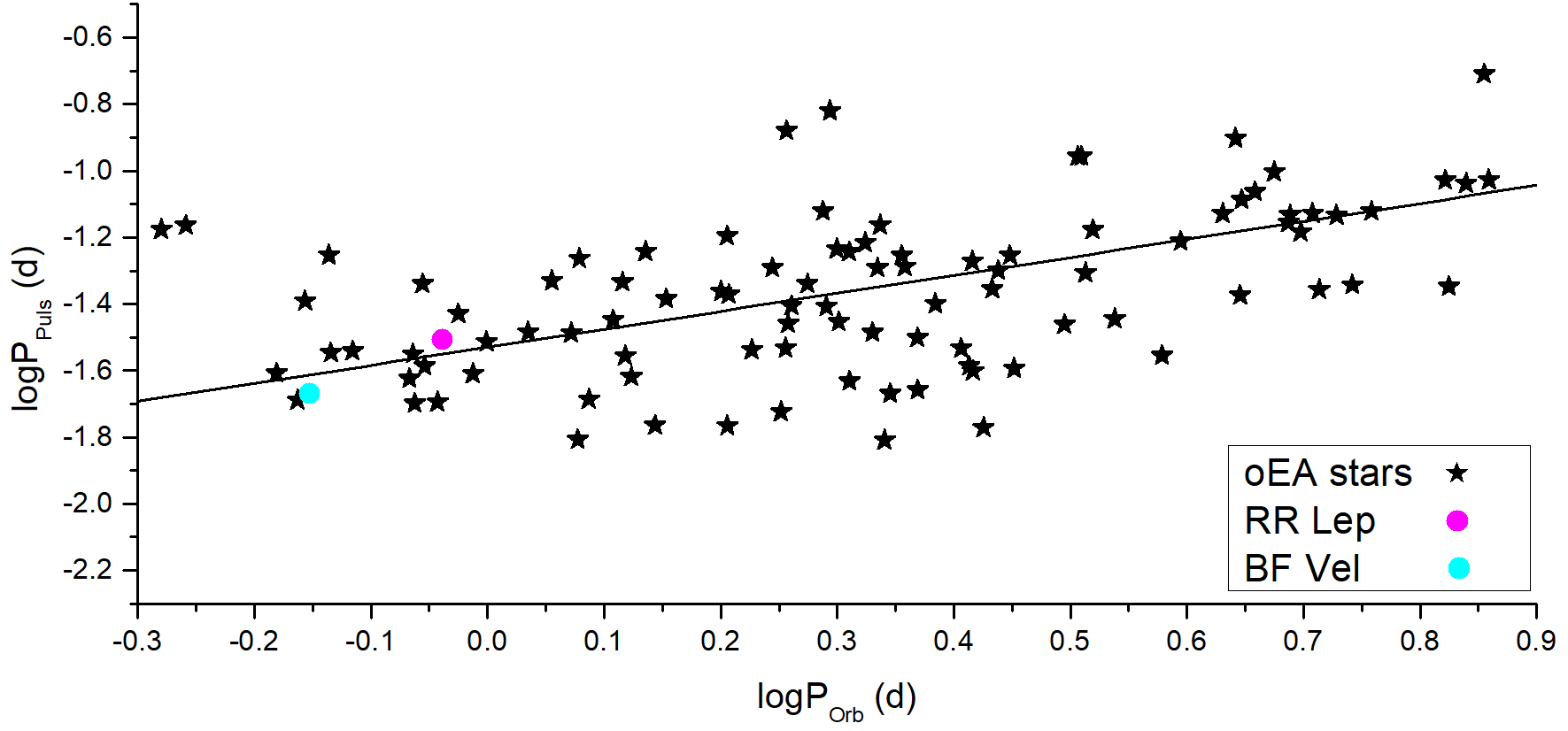}
\caption{$P_{\rm orb}-P_{\rm puls}$ correlation for $\delta$~Scuti stars of oEA systems (star symbols) and the locations of the pulsating components of RR~Lep (magenta) and BF~Vel (cyan). Black solid line represents the linear fitting of \citet{LIAN17}.}
\label{fig:PP}
\vspace{0.3cm}
\centering
\includegraphics[width=\columnwidth]{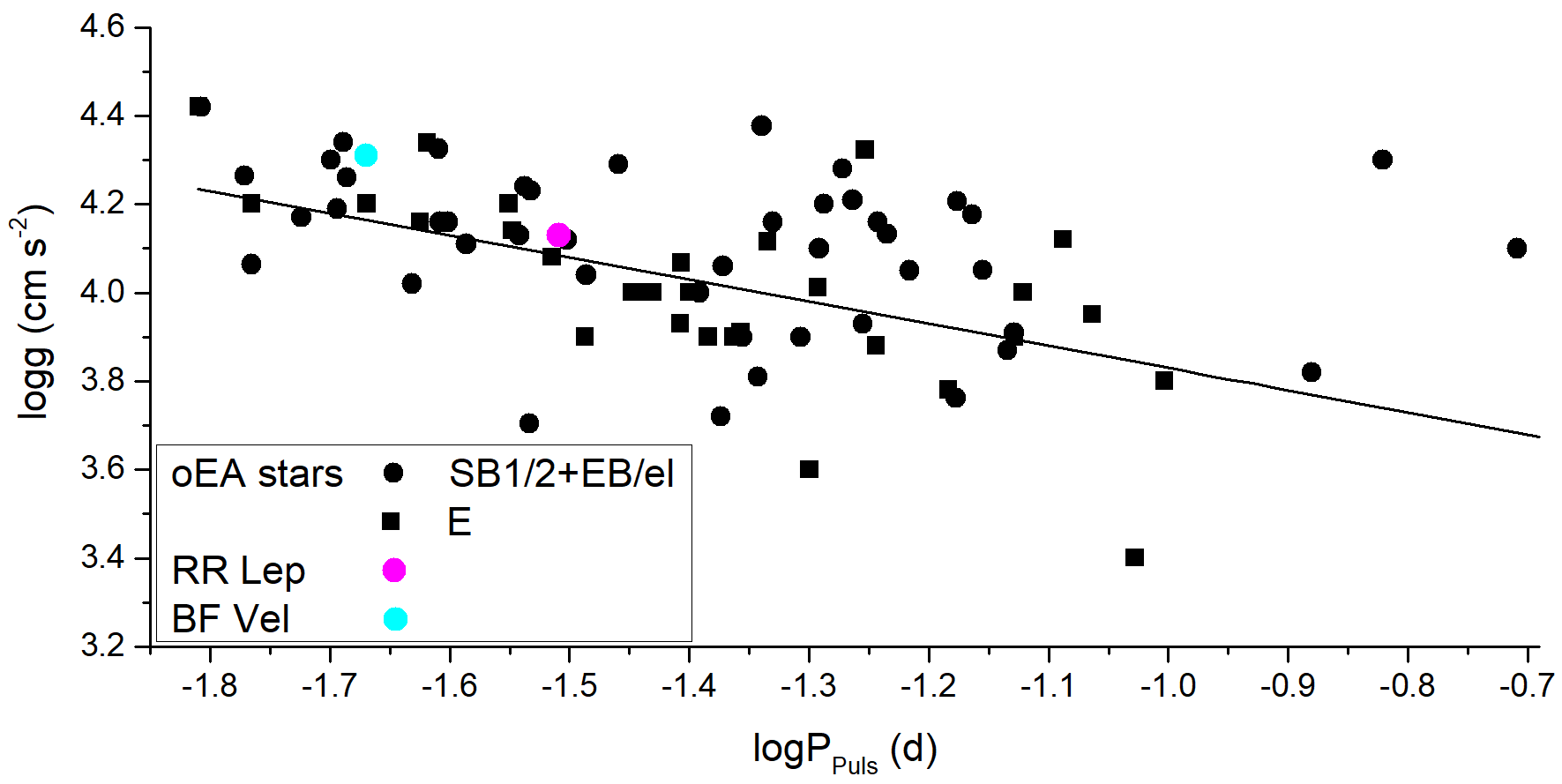}
\caption{$P_{\rm puls}-\log g$ correlation for $\delta$~Scuti stars that belong to SB1(2)+EB(el) (black dots) and EB (black squares) oEA systems and the locations of the pulsating components of RR~Lep (magenta) and BF~Vel (cyan). Black solid line represents the linear fitting of \citet{LIAN17}.}
\label{fig:GP}
\end{figure}

In order to compare the absolute and pulsation properties of the oscillating components of BF~Vel and RR~Lep with similar systems, we placed them among other $\delta$~Scuti stars of other oEA systems in two diagrams. The first diagram (Fig.~\ref{fig:PP}) is the orbital-pulsation periods ($P_{\rm orb}-P_{\rm puls}$) correlation and the second (Fig.~\ref{fig:GP}) is the pulsation period-evolutionary stage ($P_{\rm puls}-\log g$) correlation \citep[cf.][]{SOY06a, LIA12, LIAN15, LIAN17}. For both diagrams, the majority of the sample and the empirical linear fittings were taken from \citet{LIAN17}. The sample of the $P_{\rm orb}-P_{\rm puls}$ diagram includes 96 oEA systems, while that of $P_{\rm puls}-\log g$ includes 75 $\delta$~Scuti stars categorized according to the type of their host systems (i.e. SB1(2)+EB(el) or EB; for details see Sect.~\ref{Sec:AbsPar}). The reason for the different number of stars in these samples is that the absolute parameters have not been determined for every oEA system. The pulsating components of BF~Vel and RR~Lep follow the empirical trends in these plots. The primary of BF~Vel is among the ten fastest pulsators and the ten youngest stars of these samples, while its host system is among the five systems with the shortest orbital periods. The $\delta$~Scuti component of RR~Lep is, similarly to that of BF~Vel, a relatively young and fast pulsating star in comparison with the rest of these samples. These findings are in very good agreement with the main conclusions of \citet{LIA17} regarding a) the connection of the dominant frequency with the age of the star (i.e. $\log g$; the younger the star the faster its pulsation frequency) and b) the connection between the orbital and pulsations periods (i.e. the shorter the systemic orbital period the faster the pulsation frequency of the oscillating component).

Further investigations of similar systems are strongly encouraged. Expanding the study sample of oEA stars with precisely determined absolute properties will enhance our understanding of how interactions between components in close binary systems impact pulsations.

\begin{acknowledgements}
It is a pleasure to express our appreciation of the high quality and ready availability, via the Mikulski Archive for Space Telescopes (MAST), of data collected by the TESS mission. Funding for the TESS mission is provided by the NASA Explorer Program. This research partly made use of the SIMBAD database, operated at CDS, Strasbourg, France, and of NASA's Astrophysics Data System Bibliographic Services. A.~L. acknowledges financial support from the NOA's internal fellowship `SPECIES' (No.~5094). D.~J.~W.~M. and J.~F.~W. acknowledge grants for time on the ANU 2.3~m telescope from the Edward Corbould Research Fund of the Astronomical Association of Queensland. D.~J.~W.~M. is grateful to Prof.~Michael~Drinkwater for his extensive assistance with the planning and data analysis of the ANU 2.3~m telescope observations. D.~J.~W.~M. thanks C.~M.~Moriarty, A.~Mohit, Y.~Rist, L.~Coetzee, and S.~Sweet for their assistance with the spectroscopic observations. We thank the anonymous reviewer for the valuable comments that improved the quality of this work.
\end{acknowledgements}

\section*{Data Availability}

The majority of data included in this article are available as listed in the paper or from the online supplementary material it refers to. The TESS data are available online from the MAST repository (\url{https://mast.stsci.edu/portal/Mashup/Clients/Mast/Portal.html}. All the times of minima used for the O-C analysis in this work will be made available at the CDS (\url{http://cdsweb.u-strasbg.fr/}).

%
%

\bibliographystyle{aa} 
\bibliography{references.bib} 

\onecolumn
\begin{appendix}
\section{Radial velocities}
\label{Sec:RVdata}

Table~\ref{Tab:RVs} contains the RV measurements derived from our spectroscopic observations. The values in the $O-C$ columns represent the deviations of the individual measurements from the fitted RV curves. Orbital phases were calculated using the ephemeris given in Tables~\ref{Tab:Models_BFVel} and \ref{Tab:Models_RRLep}.

\begin{table*}[h!]
\begin{center}
\caption{Radial velocity measurements of the components of RR~Lep and BF~Vel.}
\label{Tab:RVs}
\scalebox{0.87}{
\begin{tabular}{cccccc|cccccc }
\hline
\hline
\multicolumn{6}{c}{RR~Lep}											&	\multicolumn{6}{c}{BF~Vel}											\\
\hline
Time        & Phase     & RV$_{1}$  & (O-C)$_{1}$   &   RV$_{2}$    &   (O-C)$_{2}$ 											&	Time        & Phase     & RV$_{1}$  & (O-C)$_{1}$   &   RV$_{2}$    &   (O-C)$_{2}$ 											\\
(HJD-2450000)   &       & (km~s$^{-1}$) & (km~s$^{-1}$) & (km~s$^{-1}$) & (km~s$^{-1}$)											&	(HJD-2450000)   &       & (km~s$^{-1}$) & (km~s$^{-1}$) & (km~s$^{-1}$) & (km~s$^{-1}$)											\\
\hline																					
58923.9740	&	0.1660	&	-106.0	&	-2.1	&	194.6	&	6.9	&	57798.0978	&	0.7439	&	137.8	&	19.9	&	-232.0	&	-11.0	\\
58923.9787	&	0.1712	&	-108.0	&	-3.1	&	208.0	&	16.1	&	57798.1016	&	0.7493	&	118.0	&	0.1	&	-232.0	&	-10.8	\\
59518.1623	&	0.2503	&	-118.0	&	-5.2	&	222.1	&	-0.2	&	57799.1614	&	0.2545	&	-112.0	&	-4.6	&	207.9	&	-23.7	\\
59518.1651	&	0.2534	&	-118.0	&	-5.2	&	207.2	&	-15.0	&	57799.1662	&	0.2615	&	-128.1	&	-21.0	&	213.0	&	-18.1	\\
59518.1679	&	0.2564	&	-120.6	&	-7.9	&	231.5	&	9.6	&	57853.0372	&	0.7795	&	117.9	&	2.0	&	-210.5	&	6.7	\\
59518.1708	&	0.2596	&	-118.0	&	-5.4	&	211.1	&	-10.5	&	57853.0403	&	0.7839	&	115.5	&	0.1	&	-210.0	&	6.0	\\
59184.9651	&	0.2696	&	-121.2	&	-9.0	&	222.0	&	2.1	&	57853.0435	&	0.7884	&	107.9	&	-6.8	&	-214.4	&	0.2	\\
59184.9682	&	0.2730	&	-118.0	&	-6.0	&	222.8	&	3.8	&	58240.8914	&	0.6853	&	113.8	&	5.0	&	-211.4	&	-8.6	\\
59184.9713	&	0.2764	&	-116.0	&	-4.3	&	198.5	&	-19.6	&	58240.8952	&	0.6907	&	110.9	&	0.7	&	-214.0	&	-8.3	\\
59184.9744	&	0.2798	&	-116.0	&	-4.6	&	234.0	&	17.0	&	58240.8990	&	0.6961	&	132.1	&	20.5	&	-218.0	&	-9.6	\\
59184.9769	&	0.2825	&	-116.0	&	-4.8	&	234.0	&	17.9	&	58240.9623	&	0.7861	&	116.0	&	1.0	&	-204.5	&	10.9	\\
59184.9793	&	0.2851	&	-116.0	&	-5.1	&	209.3	&	-5.8	&	58270.8966	&	0.3045	&	-130.4	&	-29.5	&	232.0	&	13.5	\\
59156.0223	&	0.6529	&	20.3	&	7.4	&	-246.0	&	18.0	&	58270.9625	&	0.3981	&	-99.6	&	-37.6	&	160.0	&	19.6	\\
59156.0296	&	0.6608	&	19.3	&	4.5	&	-274.2	&	-3.0	&	58271.8827	&	0.7052	&	114.0	&	0.5	&	-194.5	&	17.8	\\
59156.0369	&	0.6688	&	14.5	&	-2.0	&	-302.0	&	-24.1	&	58924.1616	&	0.1984	&	-108.0	&	-6.4	&	228.0	&	8.1	\\
59156.0436	&	0.6761	&	16.6	&	-1.4	&	-272.6	&	10.9	&	58924.1668	&	0.2058	&	-83.1	&	20.0	&	221.9	&	-1.1	\\
59156.0495	&	0.6826	&	25.0	&	5.8	&	-272.0	&	16.0	&	58924.9009	&	0.2485	&	-82.0	&	25.4	&	234.9	&	3.2	\\
59156.0554	&	0.6890	&	17.3	&	-2.9	&	-302.0	&	-9.9	&	59156.1607	&	0.7285	&	118.0	&	1.1	&	-199.7	&	19.4	\\
59157.0409	&	0.7655	&	22.0	&	-2.5	&	-302.1	&	6.5	&	59185.0595	&	0.7762	&	140.0	&	23.6	&	-202.0	&	16.1	\\
59157.0461	&	0.7712	&	22.0	&	-2.2	&	-295.8	&	11.5	&	59301.9320	&	0.7812	&	103.3	&	-12.4	&	-215.6	&	1.3	\\
59157.0513	&	0.7769	&	20.7	&	-3.1	&	-295.4	&	10.5	&	59330.0169	&	0.6730	&	97.9	&	-7.2	&	-213.1	&	-17.8	\\
	&		&		&		&		&		&	59330.0206	&	0.6782	&	105.9	&	-0.8	&	-210.8	&	-12.2	\\
	&		&		&		&		&		&	59358.9370	&	0.7508	&	114.0	&	-3.9	&	-204.2	&	17.0	\\
	&		&		&		&		&		&	59358.9401	&	0.7553	&	104.5	&	-13.4	&	-216.9	&	4.1	\\
 \hline
\end{tabular}}
\end{center}
\end{table*}

\newpage
\section{Times of minima}
\label{Sec:ToM}

Table~\ref{Tab:ToM} includes the ToM used for the ETV analyses of RR~Lep and BF~Vel (Sect.~\ref{Sec:ETV}). Particularly, we list: the HJD of the ToM, the type of observation ($Obs.~type$) based on which the respective data for its calculation were obtained (i.e. vis=visual, pg=photographic, pe=photoelectric, CCD; TESS; ASAS=CCD), and the literature source (e.g. name(s) of observer(s), campaign, mission).

\begin{table}[h]											
\begin{center}											
\caption{Times of minima (ToM) of RR~Lep and BF~Vel used in the ETV analyses.}			
\label{Tab:ToM}	
\begin{tabular}{lcc|lcc}											
\hline	
\hline										
ToM~ (HJD)	&	Obs. type	&	Source	&	ToM~ (HJD)	&	Obs. type	&	Source	\\
\hline											
\multicolumn{6}{c}{RR~Lep}											\\
\hline											
2426684.31600	&	vis	&	 Kreiner,~J.~M.	&	2446797.18700	&	pe	&	Vyas, M. L., Abhyankar, K. D.	\\
2427386.45000	&	vis	&	 Tsesevitch,~V.~P.                	&	2446814.57900	&	vis	&	 Baldwin,~M.~E., Samolyk,~G.        	\\
2430020.15400	&	pg	&	Erleksova,~G.~E.	&	2446819.15790	&	pe	&	Vyas, M. L., Abhyankar, K. D.	\\
2430377.16000	&	 pg 	&	Erleksova,~G.~E.	&	2446835.63700	&	vis	&	 Baldwin,~M.~E., Samolyk,~G.        	\\
2431057.30800	&	pg	&	Erleksova,~G.~E.	&	2446857.61100	&	vis	&	 Baldwin,~M.~E., Samolyk,~G.        	\\
2431850.07400	&	 pg 	&	Erleksova,~G.~E.	&	2447170.67800	&	vis	&	 Baldwin,~M.~E., Samolyk,~G.        	\\
2432227.22300	&	vis	&	Soloviev, A.	&	2447472.77360	&	pe	&	 Mullis,~C.~R., Faulkner,~D.~R.      	\\
2432239.10600	&	vis	&	Soloviev, A.	&	2447539.61000	&	vis	&	 Baldwin,~M.~E., Samolyk,~G.        	\\
2432260.20700	&	vis	&	Soloviev, A.	&	2447564.34000	&	vis	&	Paschke, A.	\\
2432261.12700	&	vis	&	Soloviev, A.	&	2447950.62200	&	vis	&	 Baldwin,~M.~E., Samolyk,~G.        	\\
2432553.11400	&	vis	&	Soloviev, A.	&	2448219.76600	&	vis	&	 Baldwin,~M.~E., Samolyk,~G.        	\\
2432586.12400	&	vis	&	Soloviev, A.	&	2448976.82800	&	vis	&	 Baldwin,~M.~E., Samolyk,~G.        	\\
2432595.21700	&	vis	&	Soloviev, A.	&	2450096.37300	&	vis	&	Kohl, M.	\\
2432662.97000	&	 pg 	&	Erleksova,~G.~E.	&	2450464.38800	&	vis	&	Kohl, M.	\\
2432909.20600	&	pg	&	Erleksova,~G.~E.	&	2451427.40800	&	CCD	&	 Paschke,~A. (ROTSE)             	\\
2432941.25700	&	pg	&	Erleksova,~G.~E.	&	2451510.73400	&	vis	&	 Baldwin,~M.~E., Samolyk,~G.        	\\
2432943.14500	&	vis	&	Soloviev, A.	&	2451576.63400	&	vis	&	 Baldwin,~M.~E., Samolyk,~G.        	\\
2432974.18200	&	pg	&	Erleksova,~G.~E.	&	2451870.47000	&	CCD	&	 Pojmanski,~G.             	\\
2433276.30100	&	pg	&	Erleksova,~G.~E.	&	2451908.00660	&	CCD	&	 Nagai, K.                       	\\
2433601.30800	&	pg	&	Erleksova,~G.~E.	&	2451913.07170	&	CCD	&	 Nagai, K.                       	\\
2434090.12700	&	vis	&	Soloviev, A.	&	2452170.73200	&	CCD	&	 Kosiek,~P., Ogloza,~W.           	\\
2434413.28000	&	vis	&	Erleksova,~G.~E.	&	2452228.85670	&	CCD	&	 Kosiek,~P., Ogloza,~W.           	\\
2434425.17000	&	vis	&	Erleksova,~G.~E.	&	2452312.64700	&	vis	&	 Baldwin,~M.~E., Samolyk,~G.        	\\
2434782.19000	&	vis	&	Erleksova,~G.~E.	&	2452672.38890	&	CCD	&	 Zejda, M.                       	\\
2435477.01100	&	pg	&	Erleksova,~G.~E.	&	2452681.54280	&	CCD	&	 Baldwin,~M.~E., Samolyk,~G.        	\\
2436750.37200	&	 pg 	&	Erleksova,~G.~E.	&	2453016.59200	&	CCD	&	 Dvorak, S.~W.                    	\\
2440926.55700	&	vis	&	Locher, K.	&	2453028.03170	&	CCD	&	 Kosiek,P., Ogloza,W.           	\\
2440994.30000	&	vis	&	Locher, K.	&	2453029.40630	&	CCD	&	 Zejda, M.                       	\\
2441250.62500	&	vis	&	Locher, K.	&	2453066.02010	&	CCD	&	 Kosiek, P., Ogloza, W.           	\\
2441596.62000	&	vis	&	Locher, K.	&	2453305.86320	&	CCD	&	 Baldwin,~M.~E., Samolyk,~G.        	\\
2443098.85200	&	vis	&	 Baldwin,~M.~E., Samolyk,~G.        	&	2453331.05140	&	CCD	&	 Nagai,~K.       	\\
2443143.71000	&	vis	&	 Baldwin,~M.~E., Samolyk,~G.        	&	2453406.55830	&	CCD	&	 Baldwin,~M.~E., Samolyk,~G.        	\\
2443154.70300	&	vis	&	 Baldwin,~M.~E., Samolyk,~G.        	&	2453409.30620	&	CCD	&	 Zejda, M., Mikulasek, Z.	\\
2444191.88500	&	vis	&	 Baldwin,~M.~E., Samolyk,~G.        	&	2453438.59780	&	CCD	&	 Baldwin,~M.~E., Samolyk,~G.        	\\
2444226.66350	&	pe	&	 Bookmyer,~B.~B. et al.           	&	2453737.02760	&	CCD	&	 Nagai,~K.       	\\
2444231.69680	&	pe	&	 Bookmyer,~B.~B. et al.           	&	2454043.69780	&	CCD	&	Ogloza, W.	\\
2444237.66000	&	vis	&	 Baldwin,~M.~E., Samolyk,~G.        	&	2454085.80800	&	vis	&	 Baldwin,~M.~E., Samolyk,~G.        	\\
2444291.64800	&	vis	&	 Baldwin,~M.~E., Samolyk,~G.        	&	2454108.69060	&	CCD	&	 Baldwin,~M.~E., Samolyk,~G.        	\\
2444593.74900	&	vis	&	 Baldwin,~M.~E., Samolyk,~G.        	&	2454454.71860	&	CCD	&	 Samolyk, G.                     	\\
2444616.63900	&	vis	&	 Baldwin,~M.~E., Samolyk,~G.        	&	2454465.70390	&	CCD	&	 Samolyk, G.                     	\\
2445352.65600	&	vis	&	 Baldwin,~M.~E., Samolyk,~G.        	&	2454487.67390	&	CCD	&	 Samolyk, G.                     	\\
2445373.70800	&	vis	&	 Baldwin,~M.~E., Samolyk,~G.        	&	2454520.62710	&	CCD	&	 Samolyk, G.                     	\\
2445715.15340	&	pe	&	Vyas, M. L., Abhyankar, K. D.	&	2454777.86070	&	CCD	&	 Diethelm, R.                    	\\
2445753.59200	&	vis	&	 Baldwin,~M.~E., Samolyk,~G.        	&	2454792.09200	&	CCD	&	 Nagai,~K.       	\\
2446027.31500	&	pe	&	Vyas, M. L., Abhyankar, K. D.	&	2454856.58880	&	CCD	&	 Samolyk,~G.                     	\\
2446049.28240	&	pe	&	Vyas, M. L., Abhyankar, K. D.	&	2454862.99790	&	CCD	&	 Nagai,~K.                       	\\
2446050.19900	&	pe	&	Vyas, M. L., Abhyankar, K. D.	&	2455144.94770	&	CCD	&	 Diethelm, R.                    	\\
2446445.66700	&	vis	&	 Baldwin,~M.~E., Samolyk,~G.        	&	2455157.76390	&	CCD	&	 Samolyk,~G.                     	\\
2446489.60500	&	vis	&	 Baldwin,~M.~E., Samolyk,~G.        	&	2455513.85860	&	CCD	&	 Diethelm, R.                    	\\
2446785.28510	&	pe	&	Vyas, M. L., Abhyankar, K. D.	&	2455953.26740	&	CCD	&	 Liakos, A. et al.               	\\
2446795.38500	&	vis	&	 BBSAG observers                	&	2455959.67420	&	CCD	&	 Samolyk,~G.                     	\\
\hline					
\end{tabular}
\end{center}					
\end{table}

\begin{table}											
\begin{center}											
\caption*{Table~\ref{Tab:ToM} (continued)}			
\begin{tabular}{lcc|lcc}											
\hline	
\hline										
ToM~ (HJD)	&	Obs. type	&	Source	&	ToM~ (HJD)	&	Obs. type	&	Source	\\
\hline	
\hline											
2456238.88040	&	CCD	&	 Diethelm, R.                    	&	2458494.48940	&	CCD	&	 Samolyk,~G.                     	\\
2456302.04583	&	CCD	&	 Nagai,~K.       	&	2458495.40120	&	CCD	&	 Samolyk,~G.                     	\\
2456651.73730	&	CCD	&	 Samolyk,~G.                     	&	2458792.91790	&	CCD	&	 Samolyk,~G.                     	\\
2457320.91280	&	CCD	&	 Samolyk,~G.                     	&	2458868.89465	&	CCD	&	 Nagai,~K.       	\\
2457356.61560	&	CCD	&	 Samolyk,~G.                     	&	2458885.37560	&	CCD	&	 Samolyk,~G.                     	\\
2457421.61080	&	CCD	&	 Samolyk,~G.                     	&	2458896.35960	&	CCD	&	 Samolyk,~G.                     	\\
2457699.89770	&	CCD	&	 Samolyk,~G.                     	&	2459174.64868	&	TESS	&	This study	\\
2457758.95063	&	CCD	&	 Nagai,~K.       	&	2459175.10678	&	TESS	&	This study	\\
2458093.07450	&	CCD	&	 Nagai,~K.       	&	2459184.71818	&	TESS	&	This study	\\
2458123.73930	&	CCD	&	 Samolyk,~G.                     	&	2459185.17528	&	TESS	&	This study	\\
2458438.64647	&	TESS	&	This study	&	2459187.46458	&	TESS	&	This study	\\
2458439.10347	&	TESS	&	This study	&	2459187.92158	&	TESS	&	This study	\\
2458449.17357	&	TESS	&	This study	&	2459192.04280	&	CCD	&	 Richards, T. et al.             	\\
2458449.63167	&	TESS	&	This study	&	2459198.90778	&	TESS	&	This study	\\
2458450.08900	&	CCD	&	 Nagai,~K.       	&	2459199.36518	&	TESS	&	This study	\\
2458452.37817	&	TESS	&	This study	&	2459518.84150	&	CCD	&	 Samolyk,~G.                     	\\
2458452.83417	&	TESS	&	This study	&	2459610.38910	&	CCD	&	 Samolyk,~G.                     	\\
2458462.44807	&	TESS	&	This study	&	2459843.82720	&	CCD	&	Hazel, L.	\\
2458462.90497	&	TESS	&	This study	&	2459875.86370	&	CCD	&	 Samolyk,~G.                     	\\
2458470.68610	&	CCD	&	 Samolyk,~G.                     	&	2459956.42200	&	CCD	&	Paschke, A.	\\
\hline											
\multicolumn{6}{c}{BF~Vel}											\\
\hline											
2428102.40200	&	 pg 	&	 O'Leary, W., O'Connell, D.        	&	2459264.06796	&	ASAS-SN	&	This study	\\
2428253.07000	&	 pg 	&	 O'Leary, W.                     	&	2459264.77259	&	TESS	&	This study	\\
2451868.26900	&	ASAS	&	 Paschke, A.               	&	2459265.82950	&	TESS	&	This study	\\
2452235.77471	&	ASAS-3	&	This study	&	2459266.88456	&	TESS	&	This study	\\
2452296.66870	&	ASAS	&	 Zakrzewski, B.            	&	2459278.85380	&	TESS	&	This study	\\
2452297.02330	&	ASAS	&	 Zakrzewski, B.            	&	2459279.20590	&	TESS	&	This study	\\
2452500.48700	&	CCD	&	Kreiner, B. Nelson SS	&	2459305.60627	&	CCD	&	This study	\\
2452699.73200	&	ASAS	&	 Dvorak, S.                 	&	2459307.71803	&	CCD	&	This study	\\
2453307.65500	&	ASAS	&	 Zakrzewski, B.            	&	2459628.05348	&	ASAS-SN	&	This study	\\
2453308.00240	&	ASAS	&	 Zakrzewski, B.            	&	2459923.04011	&	ASAS-SN	&	This study	\\
2453463.59225	&	ASAS-3	&	This study	&	2459937.82490	&	CCD	&	This study	\\
2453564.97360	&	INTEGRAL	&	 Zakrzewski, B.        	&	2459940.64090	&	CCD	&	This study	\\
2453656.85870	&	INTEGRAL	&	 Zakrzewski, B.        	&	2459941.69680	&	CCD	&	This study	\\
2453748.37070	&	CCD	&	 Manimanis et al.	&	2459942.75300	&	CCD	&	This study	\\
2453749.42710	&	CCD	&	 Manimanis et al.	&	2459943.80850	&	CCD	&	This study	\\
2454533.70961	&	ASAS-3	&	This study	&	2459946.62598	&	CCD	&	This study	\\
2454570.67300	&	ASAS-3	&	 Zakrzewski, B.            	&	2459947.68207	&	CCD	&	This study	\\
2454631.56910	&	ASAS-3	&	 Zakrzewski, B.            	&	2459948.73830	&	CCD	&	This study	\\
2456959.78410	&	CCD	&	 Jurysek, J. et al.              	&	2459949.79330	&	CCD	&	This study	\\
2457175.92110	&	CCD	&	 Pavlov, H. et al.               	&	2459952.60960	&	CCD	&	This study	\\
2457764.48748	&	ASAS-SN	&	This study	&	2459990.62850	&	TESS	&	This study	\\
2457777.16064	&	CCD	&	This study	&	2459990.98120	&	TESS	&	This study	\\
2457850.02680	&	ASAS-SN	&	 Zakrzewski, B.            	&	2459996.96490	&	TESS	&	This study	\\
2457877.83500	&	ASAS-SN	&	 Zakrzewski, B.            	&	2459997.31760	&	TESS	&	This study	\\
2458181.27215	&	ASAS-SN	&	This study	&	2460001.18830	&	TESS	&	This study	\\
2458527.65360	&	TESS	&	This study	&	2460001.54050	&	TESS	&	This study	\\
2458532.58068	&	ASAS-SN	&	This study	&	2460005.41330	&	TESS	&	This study	\\
2458555.11100	&	TESS	&	This study	&	2460005.76590	&	TESS	&	This study	\\
2458860.65741	&	ASAS-SN	&	This study	&	2460013.15780	&	TESS	&	This study	\\
2459256.32500	&	TESS	&	This study	&	2460013.51040	&	TESS	&	This study	\\
2459256.67700	&	TESS	&	This study	&						\\
\hline	
\end{tabular}
\end{center}
\end{table}


\section{Combination pulsation frequencies}
\label{Sec:PulsMod}

Table~\ref{Tab:FreqsDep} contains the increasing number of the frequency ($n$), its value ($f_{\rm n}$), amplitude ($A$), phase ($\Phi$), S/N, and the most possible frequency combination ($Combo$). This table is complementary to Table~\ref{Tab:FreqsInd} and their integration is the complete model of the pulsational behaviour of each studied system (Sect.~\ref{Sec:Puls}).
\begin{sidewaystable*}
\begin{center}											
\caption{Combination pulsation frequencies of the primary components of RR~Lep and BF~Vel.}	
\label{Tab:FreqsDep}	
\begin{tabular}{cccccc|cccccc}														
\hline		
\hline																				
$n$	&	  $f_{\rm n}$	&	$A$	&	  $\Phi$	&	S/N	&	Combo	&	$n$	&	  $f_{\rm n}$	&	$A$	&	  $\Phi$	&	S/N	&	Combo	\\
	&	     (d$^{-1}$)	&	(mmag)	&	($2\pi$~rad)	&		&		&		&	     (d$^{-1}$)	&	(mmag)	&	($2\pi$~rad)	&		&		\\
\hline																
\multicolumn{6}{c}{RR~Lep}											&	\multicolumn{6}{c}{BF~Vel}											\\
\hline																							
3	&	29.6794$\pm$0.0001	&	2.78$\pm$0.01	&	0.589$\pm$0.001	&	32.6	&	$f_2-2f_{\rm orb}$	&	1	&	1.4205$\pm$0.0003	&	1.81$\pm$0.02	&	0.001$\pm$0.002	&	38.59	&	$f_{\rm orb}$	\\
4	&	30.0925$\pm$0.0001	&	2.08$\pm$0.01	&	0.418$\pm$0.001	&	21.5	&	$f_1-2f_{\rm orb}$	&	2	&	2.8408$\pm$0.0005	&	1.12$\pm$0.02	&	0.127$\pm$0.003	&	31.24	&	2$f_{\rm orb}$	\\
8	&	25.7836$\pm$0.0003	&	0.94$\pm$0.01	&	0.799$\pm$0.002	&	15.5	&	$f_5+2f_{\rm orb}$	&	4	&	7.1017$\pm$0.0007	&	0.77$\pm$0.02	&	0.858$\pm$0.005	&	26.85	&	5$f_{\rm orb}$	\\
9	&	26.4520$\pm$0.0003	&	0.88$\pm$0.01	&	0.326$\pm$0.003	&	14.3	&	$f_5+f_7-f_2+2f_{\rm orb}$	&	5	&	9.9401$\pm$0.0006	&	0.79$\pm$0.02	&	0.125$\pm$0.005	&	33.01	&	7$f_{\rm orb}$	\\
10$^a$	&	0.0273$\pm$0.0003	&	0.87$\pm$0.01	&	0.542$\pm$0.003	&	12.1	&	$f_6-f_1+2f_{\rm orb}$	&	6$^a$	&	0.0351$\pm$0.0008	&	0.64$\pm$0.02	&	0.058$\pm$0.006	&	11.52	&	$\sim 2f_{\rm orb}-2f_{\rm orb}$	\\
11	&	34.6782$\pm$0.0004	&	0.80$\pm$0.01	&	0.728$\pm$0.003	&	11.3	&	$f_7+2f_{\rm orb}$	&	7	&	43.8929$\pm$0.0009	&	0.59$\pm$0.02	&	0.009$\pm$0.006	&	15.68	&	$f_3-2f_{\rm orb}$	\\
12	&	27.9663$\pm$0.0004	&	0.80$\pm$0.01	&	0.407$\pm$0.003	&	12.4	&	$f_1-4f_{\rm orb}$	&	8	&	49.6430$\pm$0.0009	&	0.57$\pm$0.02	&	0.215$\pm$0.006	&	11.39	&	$\sim f_3+2f_{\rm orb}$	\\
13	&	33.3627$\pm$0.0004	&	0.76$\pm$0.01	&	0.439$\pm$0.003	&	10.0	&	$f_1+f_{\rm orb}$	&	9$^a$	&	35.509$\pm$0.001	&	0.54$\pm$0.02	&	0.856$\pm$0.007	&	14.56	&	$f_3+2f_6-8f_{\rm orb}$	\\
14	&	24.2673$\pm$0.0004	&	0.70$\pm$0.01	&	0.749$\pm$0.003	&	12.8	&	$f_3+f_5+f_7-2f_2+2f_{\rm orb}$	&	10	&	12.780$\pm$0.001	&	0.42$\pm$0.02	&	0.372$\pm$0.009	&	14.61	&	9$f_{\rm orb}$	\\
15	&	32.8930$\pm$0.0005	&	0.56$\pm$0.01	&	0.767$\pm$0.004	&	6.5	&	$f_2+f_{\rm orb}$	&	11	&	17.046$\pm$0.001	&	0.43$\pm$0.02	&	0.672$\pm$0.008	&	13.79	&	12$f_{\rm orb}$	\\
16	&	31.8601$\pm$0.0006	&	0.51$\pm$0.01	&	0.191$\pm$0.004	&	6.3	&	$\sim f_2$	&	12	&	43.963$\pm$0.001	&	0.42$\pm$0.02	&	0.221$\pm$0.009	&	11.05	&	$\sim f_3-2f_{\rm orb}$	\\
17	&	1.0680$\pm$0.0006	&	0.50$\pm$0.01	&	0.847$\pm$0.004	&	8.1	&	$f_{\rm orb}$	&	13$^a$	&	0.174$\pm$0.001	&	0.36$\pm$0.02	&	0.15$\pm$0.01	&	6.79	&	$\sim2f_6$	\\
18	&	32.1310$\pm$0.0006	&	0.48$\pm$0.01	&	0.301$\pm$0.005	&	5.9	&	$f_2+f_7-f_1$ 	&	14	&	42.457$\pm$0.001	&	0.35$\pm$0.02	&	0.63$\pm$0.01	&	10.31	&	$f_3-3f_{\rm orb}$	\\
19	&	5.4627$\pm$0.0007	&	0.45$\pm$0.01	&	0.222$\pm$0.005	&	18.1	&	5$f_{\rm orb}$	&	15$^a$	&	46.505$\pm$0.002	&	0.32$\pm$0.02	&	0.92$\pm$0.01	&	6.59	&	$f_3-2f_6$	\\
20	&	27.6038$\pm$0.0007	&	0.41$\pm$0.01	&	0.615$\pm$0.005	&	6.7	&	$f_2+f_6-f_1-2f_{\rm orb}$	&	16$^a$	&	0.122$\pm$0.002	&	0.31$\pm$0.02	&	0.04$\pm$0.01	&	5.69	&	$\sim2f_6$	\\
21	&	31.9868$\pm$0.0007	&	0.43$\pm$0.01	&	0.661$\pm$0.005	&	5.3	&	$2f_1-f_7$	&	17	&	14.200$\pm$0.002	&	0.30$\pm$0.02	&	0.55$\pm$0.01	&	10.68	&	10$f_{\rm orb}$	\\
22	&	33.0801$\pm$0.0006	&	0.47$\pm$0.01	&	0.980$\pm$0.005	&	5.7	&	$2f_1-f_7+f_{\rm orb}$	&	18	&	36.953$\pm$0.002	&	0.27$\pm$0.02	&	0.17$\pm$0.01	&	7.30	&	$26f_{\rm orb}$	\\
23	&	31.1839$\pm$0.0007	&	0.40$\pm$0.01	&	0.209$\pm$0.005	&	4.8	&	$f_1-f_{\rm orb}$	&	19	&	11.363$\pm$0.002	&	0.26$\pm$0.02	&	0.62$\pm$0.01	&	8.96	&	8$f_{\rm orb}$	\\
24	&	28.9895$\pm$0.0007	&	0.40$\pm$0.01	&	0.871$\pm$0.006	&	5.0	&	$f_1-3f_{\rm orb}$	&	20	&	8.523$\pm$0.002	&	0.26$\pm$0.02	&	0.43$\pm$0.01	&	9.99	&	6$f_{\rm orb}$	\\
25	&	13.1081$\pm$0.0009	&	0.35$\pm$0.01	&	0.778$\pm$0.006	&	13.0	&	12$f_{\rm orb}$	&	21$^a$	&	34.582$\pm$0.002	&	0.25$\pm$0.02	&	0.84$\pm$0.01	&	7.15	&	$f_3-9f_{\rm orb}+19f_{6}$	\\
26	&	15.2928$\pm$0.0008	&	0.37$\pm$0.01	&	0.358$\pm$0.006	&	14.6	&	14$f_{\rm orb}$	&	22	&	19.888$\pm$0.002	&	0.24$\pm$0.02	&	0.85$\pm$0.01	&	9.47	&	14$f_{\rm orb}$	\\
27	&	30.7746$\pm$0.0009	&	0.34$\pm$0.01	&	0.731$\pm$0.007	&	5.0	&	$f_2-f_{\rm orb}$	&	23$^a$	&	47.618$\pm$0.002	&	0.23$\pm$0.02	&	0.61$\pm$0.02	&	5.32	&	$2f_3-32f_{\rm orb}-4f_{6}$	\\
28	&	29.7359$\pm$0.0009	&	0.33$\pm$0.01	&	0.303$\pm$0.007	&	4.8	&	$\sim f_3$	&	24$^a$	&	43.305$\pm$0.002	&	0.21$\pm$0.02	&	0.59$\pm$0.02	&	6.18	&	$2f_3-35f_{\rm orb}-4f_{6}$	\\
29	&	7.645$\pm$0.001	&	0.31$\pm$0.01	&	0.289$\pm$0.007	&	13.7	&	7$f_{\rm orb}$	&	25$^a$	&	0.381$\pm$0.002	&	0.21$\pm$0.02	&	0.04$\pm$0.02	&	4.05	&	$4f_{6}$	\\
30	&	27.477$\pm$0.001	&	0.29$\pm$0.01	&	0.195$\pm$0.007	&	5.1	&	$f_2-4f_{\rm orb}$	&	26	&	39.806$\pm$0.003	&	0.21$\pm$0.02	&	0.29$\pm$0.02	&	4.72	&	28$f_{\rm orb}$	\\
31	&	31.435$\pm$0.001	&	0.28$\pm$0.01	&	0.360$\pm$0.008	&	4.8	&	$f_7-f_{\rm orb}$	&	27$^a$	&	47.776$\pm$0.003	&	0.20$\pm$0.02	&	0.78$\pm$0.02	&	4.91	&	$2f_3-32f_{\rm orb}-2f_{6}$	\\
32	&	22.578$\pm$0.001	&	0.28$\pm$0.01	&	0.280$\pm$0.008	&	6.3	&	$f_5-f_{\rm orb}$ 	&	28	&	4.276$\pm$0.003	&	0.20$\pm$0.02	&	0.65$\pm$0.02	&	5.54	&	3$f_{\rm orb}$	\\
33	&	25.415$\pm$0.001	&	0.28$\pm$0.01	&	0.495$\pm$0.008	&	4.8	&	$f_5+f_7-f_2+f_{\rm orb}$	&	29$^a$	&	0.206$\pm$0.003	&	0.20$\pm$0.02	&	0.07$\pm$0.02	&	3.73	&	$\sim2f_6$	\\
34$^a$	&	0.062$\pm$0.001	&	0.28$\pm$0.01	&	0.524$\pm$0.008	&	4.8	&	$\sim 2f_{10}$	&	30	&	48.221$\pm$0.003	&	0.19$\pm$0.02	&	0.07$\pm$0.02	&	4.65	&	$\sim f_3+f_{\rm orb}$	\\
35	&	27.559$\pm$0.001	&	0.27$\pm$0.01	&	0.868$\pm$0.008	&	4.9	&	$\sim f_{20}$	&	31$^a$	&	41.377$\pm$0.003	&	0.19$\pm$0.02	&	0.75$\pm$0.02	&	4.89	&	$f_3-4f_{\rm orb}+4f_{6}$	\\
36	&	2.189$\pm$0.001	&	0.26$\pm$0.01	&	0.623$\pm$0.008	&	6.4	&	$2f_{\rm orb}$	&	32$^a$	&	37.620$\pm$0.003	&	0.18$\pm$0.02	&	0.60$\pm$0.02	&	4.88	&	$2f_3-39f_{\rm orb}-4f_{6}$	\\
	&		&		&		&		&		&	33$^a$	&	0.527$\pm$0.003	&	0.18$\pm$0.02	&	0.30$\pm$0.02	&	3.35	&	$6f_{6}$	\\
	&		&		&		&		&		&	34	&	29.826$\pm$0.003	&	0.18$\pm$0.02	&	0.31$\pm$0.02	&	6.49	&	21$f_{\rm orb}$	\\
	&		&		&		&		&		&	35	&	53.973$\pm$0.003	&	0.18$\pm$0.02	&	0.62$\pm$0.02	&	5.87	&	38$f_{\rm orb}$	\\
	&		&		&		&		&		&	36$^a$	&	43.669$\pm$0.003	&	0.18$\pm$0.02	&	0.16$\pm$0.02	&	4.97	&	$f_3-2f_{\rm orb}-2f_6$	\\
	&		&		&		&		&		&	37	&	32.668$\pm$0.003	&	0.18$\pm$0.02	&	0.53$\pm$0.02	&	6.80	&	23$f_{\rm orb}$	\\

\hline																																						
\end{tabular}
\\$^{\rm a}$ possible leftover and not an intrinsic frequency
\end{center}
\end{sidewaystable*}
\end{appendix}
\end{document}